\pdfoutput=1
\documentclass[11pt]{article}

\usepackage[final]{acl2024}
\usepackage{times}
\usepackage{latexsym}

\usepackage[T1]{fontenc}
\usepackage{microtype}

\usepackage{inconsolata}

\usepackage[utf8]{inputenc}
\usepackage{booktabs,multirow}
\usepackage{enumitem}
\usepackage{kotex} % Menu - Tex Live version - 2021 로 설정필요
\usepackage{color}
\usepackage{rotating}
\usepackage{amsmath}
\usepackage{amsfonts}
\usepackage{pifont}
\usepackage{array}
\usepackage{amssymb}
\usepackage{algpseudocode}
\usepackage{algorithm}
\usepackage{float}
\usepackage{caption}
\usepackage{tcolorbox}
\usepackage{multicol}
\usepackage{lipsum}
\usepackage[normalem]{ulem}
\usepackage{xcolor}
\usepackage{soul}
\usepackage{listings}
\usepackage{xcolor}
\usepackage{amsmath} % For LaTeX math symbols
\usepackage{tabularx}
\usepackage{graphicx}
\usepackage{natbib}
\usepackage{makecell}
\usepackage{placeins}
\usepackage[export]{adjustbox}   % ← 여기에 추가

\definecolor{customhighlight}{HTML}{DDB6E4}
\sethlcolor{customhighlight}

\colorlet{instructionbg}{gray!15}
\colorlet{questionbg}{gray!25}

\algrenewcommand\algorithmicrequire{\textbf{Input:}}
\algrenewcommand\algorithmicensure{\textbf{Output:}}

\newcommand{\cmark}{\textcolor{green!70!black}{\ding{51}}} 
\newcommand{\xmark}{\textcolor{red}{\ding{55}}}            
\newcommand{\ie}{{\it i.e.}}

\title{Multi-view-guided Passage Reranking with Large Language Models}

\author{Jeongwoo Na\thanks{\ \ Equal contribution.}, Jun Kwon\footnotemark[1], Eunseong Choi, Jongwuk Lee\thanks{\ \ Corresponding author.} \\
        Sungkyunkwan University, Republic of Korea\\  
        \texttt{\{wjddn7946, kwon04210, eunseong, jongwuklee\}@skku.edu}}

\begin{document}
\maketitle

\begin{abstract}
Recent advances in large language models (LLMs) have shown impressive performance in \emph{passage reranking} tasks. Despite their success, LLM-based methods still face challenges in \emph{efficiency} and \emph{sensitivity to external biases}. (i) Existing models rely mostly on autoregressive generation and sliding window strategies to rank passages, which incurs heavy computational overhead as the number of passages increases. (ii) External biases, such as position or selection bias, hinder the model’s ability to accurately represent passages and the input-order sensitivity. To address these limitations, we introduce a novel passage reranking model, called \textit{\textbf{M}\textit{ulti-\textbf{V}iew-guided \textbf{P}assage Reranking} (\textbf{MVP})}. MVP is a non-generative LLM-based reranking method that encodes query–passage information into diverse view embeddings without being influenced by external biases. For each view, it combines query-aware passage embeddings to produce a distinct anchor vector, used to directly compute relevance scores in \textit{a single decoding step}. Besides, it employs an orthogonal loss to make the views more distinctive. Extensive experiments demonstrate that MVP, with just 220M parameters, matches the performance of much larger 7B-scale fine-tuned models while achieving a 100× reduction in inference latency. Notably, the 3B-parameter variant of MVP achieves state-of-the-art performance on both in-domain and out-of-domain benchmarks. The source code is available at \url{https://github.com/bulbna/MVP}.

\end{abstract}

\section{Introduction}\label{sec:introduction}

\begin{figure}[t]
\includegraphics[width=0.95\linewidth]{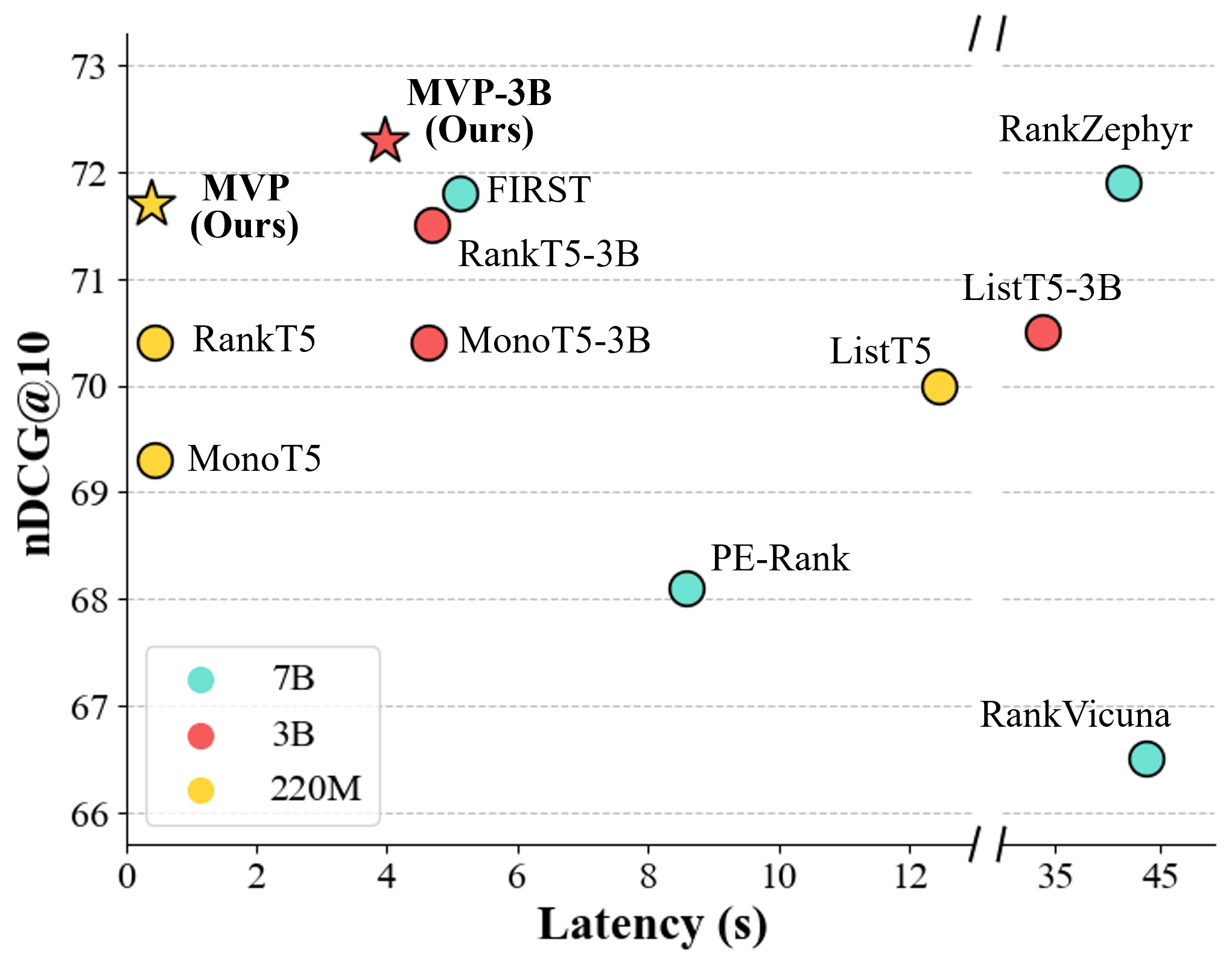}
\caption{Comparison of latency and nDCG@10 across various reranking models. Latency refers to the time required to rerank for a single query and nDCG@10 is averaged over DL19 and DL20.}\label{fig:fig_MVP_motivation}
\vskip -0.1in
\end{figure}
% ~\cite{eacl/IzacardG21/FiD}plt.savefig("fig_MVP_num_rel_trec.png", dpi=300, bbox_inches='tight')

% Task 소개 -> 최근 연구 동향

Passage reranking aims to assign fine-grained relevance scores to candidate passages -- typically retrieved by a first-stage retriever~\cite{trec/RobertsonWJHG94/BM25, emnlp/KarpukhinOMLWEC20/DPR} -- by harnessing the language understanding capabilities of large language models (LLMs), in both zero-shot and fine-tuned settings. Recent studies~\cite{emnlp/0001YMWRCYR23/RankGPT, tmlr/LiangBLTSYZNWKN23/UNSUP_HOLISTIC} formulate a prompt that consists of a query and candidate passages and generate an ordered list of passage identifiers in a zero-shot setting. Subsequent work has fine-tuned open-source LLMs by distilling knowledge from the teacher model~\cite{corr/abs-2309-15088/RankVicuna, corr/abs-2312-02724/RankZephyr}, achieving competitive performance.

Despite their success, LLM-based reranking methods still face challenges in \emph{efficiency} and \emph{sensitivity to input order}. Specifically, we address two key issues for designing an efficient and effective LLM-based reranker.

\begin{figure}[t]
\includegraphics[width=0.95\linewidth]{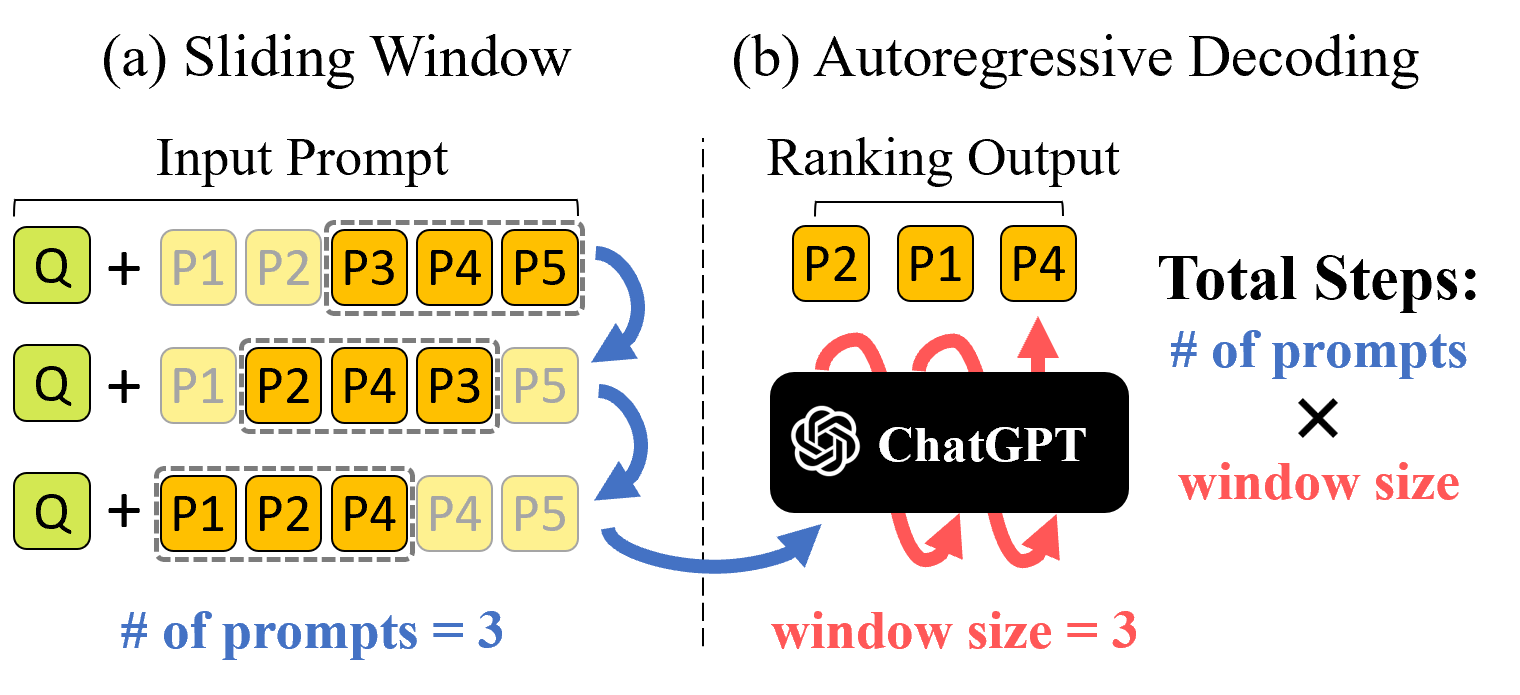}
\caption{Inference pipeline of a generative listwise reranker. The total number of inferences is determined by the product of (i) the number of prompts and (ii) the window size required to evaluate all candidate passages.}\label{fig:fig_MVP_SWAD}
\vskip -0.2in
\end{figure}

% challenge 1. Efficiency (Figure 1에 대한 설명)
\noindent(i) \textit{How do we perform reranking without incurring unnecessary inference?} 
Efficient reranking hinges on two key aspects: \textbf{global ranking} (evaluating all candidates at once) and \textbf{single pass decoding} (performing reranking with a single decoding step). However, existing methods~\cite{corr/abs-2309-15088/RankVicuna, corr/abs-2312-02724/RankZephyr} fail to satisfy both. First, they are unable to include all candidate passages in a single prompt due to input length limitations, leading to rely on sliding-window algorithms, as illustrated in Figure~\ref{fig:fig_MVP_SWAD}(a). Next, generative rerankers employ autoregressive decoding, generating one passage identifier at a time, which leads to substantial computational overhead in Figure~\ref{fig:fig_MVP_SWAD}(b). 
% \noindent(i) \textit{How do we perform reranking without incurring unnecessary inference?} 
% Efficient reranking hinges on two key aspects: \textbf{global ranking} (evaluating all candidates at once) and \textbf{single pass decoding} (performing reranking with a single decoding step). First, existing methods~\cite{corr/abs-2309-15088/RankVicuna, corr/abs-2312-02724/RankZephyr} are unable to include all candidate passages in a single prompt due to input length limitations, leading to rely on sliding-window algorithms, as illustrated in Figure~\ref{fig:fig_MVP_SWAD}(a). Next, generative rerankers employ autoregressive decoding, generating one passage identifier at a time, which leads to substantial computational overhead in Figure ~\ref{fig:fig_MVP_SWAD}(b). 

%2. explicit embedding
\noindent(ii) \textit{How do we represent query-passage explicitly without introducing bias?}
% Query-pasasge relationship를 정확하게 표현하는 것에서 common bias가 영향을 줄 수 있다.
While LLMs show strong zero-shot reranking performance, the unbiased modeling of query-passage relationships remains an underexplored challenge due to common biases~\cite{kdd/DaiXXPDX24/BIAS}. 
First, \textbf{position bias} emerges in long-context prompts, a problem known as \textit{lost-in-the-middle}~\cite{tacl/LiuLHPBPL24/Lost-in-the-middle}. Second, \textbf{selection bias} arises when natural language tokens (e.g., "A", "1") are used as passage identifiers. These identifiers may encode unintended priors, potentially biasing reranking—as observed in multiple-choice settings~\cite{iclr/Zheng0M0H24/NotRobustMC}.

% Generative listwise reranker 의 추론 파이프라인. 전체 추론횟수는 전체 candidates에 대한 평가를 위한 (i) 프롬프트의 수, (ii) window 크기의 곱으로 나타냄

%MVP performs global ranking in a single decoder pass without external biases. 

To this end, we propose a novel listwise reranking model, \textit{\textbf{M}\textit{ulti-\textbf{V}iew-guided \textbf{P}assage reranking} (\textbf{MVP})}. It consists of two key components under the Fusion-in-Decoder (FiD) architecture~\cite{corr/abs-2007-01282/FiD}.

\noindent\textbf{Multi-View Encoding.} Each query–passage pair is encoded into learnable soft prompts in the FiD architecture. To eliminate position bias, soft prompts are inserted at the same fixed positions across all passages. For each passage, distinct position embeddings are used to separate relevant views. Since these prompts are not tied to any pre-trained vocabulary, they allow for unbiased modeling of query-passage relationships. The encoder produces view-specific embeddings, called relevance vectors, which are then passed to the decoder to compute the final relevance scores. 

\noindent\textbf{Anchor-Guided Decoding.}
Our method adopts a non-generative design that leverages anchor vectors for listwise relevance scoring across all candidates within a single decoding step. This approach operates independently from a language modeling (LM) head. During decoding, MVP aggregates view-specific relevance embeddings from all candidate passages using cross-attention in the decoder to produce anchor vectors. This design directly computes similarity-based scores, aligning both training and inference with the ranking objective while substantially improving efficiency. 

%figure 1 실험 설명
As illustrated in Figure~\ref{fig:fig_MVP_motivation}, MVP-3B achieves state-of-the-art performance on in-domain benchmarks (DL19 and DL20). Notably, our 220M-parameter model matches the nDCG@10 of 7B-scale listwise rerankers while reducing inference latency by up to 100×. These results highlight the efficiency and scalability of our reranking approach, demonstrating that high-quality reranking can be achieved without the computational overhead of large scale generative models.

Our contributions are summarized as follows:
\vspace{-0.7em}

\begin{itemize}[leftmargin=0.5em, itemsep=-0.1em]
\item \textbf{Efficient Listwise Reranker}: We propose a novel non-generative reranking method named \textbf{MVP}, which enables global ranking in a single step.

\item \textbf{Robustness to External Biases}: Our embedding-based architecture is robust to position and selection biases, enabling flexible adaptation to diverse passage input scenarios. 

\item \textbf{Extensive Experiments}: MVP achieves state-of-the-art performance on both in-domain and out-domain benchmarks.
\end{itemize}
\section{Related Work} \label{sec:related_work}
\subsection{Reranking with LLMs}\label{sec:related_work-gen}
Recent work has explored leveraging the language understanding capabilities of LLMs for passage reranking~\cite{corr/abs-2308-07107/SURVEY_LLM4IR, tmlr/LiangBLTSYZNWKN23/UNSUP_HOLISTIC}. Depending on the prompting strategy, methods can be categorized into pointwise and listwise approaches.
Pointwise rerankers estimate the relevance between a query and a single passage. For example, Some pointwise approaches~\cite{corr/abs-1910-14424/monoBERT, emnlp/NogueiraJPL20/MonoT5, naacl/ZhuangQHWYWB24/UNSUP_Beyond_YN} compute relevance scores using the logits of relevance-related tokens such as “Yes” or “No”. Other approaches~\cite{emnlp/SachanLJAYPZ22/UNSUP_UPR, emnlp/Zhuang0KZ23/UNSUP_QLM, acl/ChoJSP23/Co-Prompt} estimate the relevance of a passage based on the probability of generating the corresponding query sequence.
In contrast, \citet{nips/listbetterpoint} demonstrated that listwise reranking methods outperform pointwise approaches by comparing candidate passages at once. Building on this, RankGPT~\cite{emnlp/0001YMWRCYR23/RankGPT} employed GPT-4~\cite{corr/abs-2303-08774/GPT-4} to achieve state-of-the-art zero-shot reranking performance, and later work distilled knowledge into open-source LLMs~\cite{corr/abs-2309-15088/RankVicuna, corr/abs-2312-02724/RankZephyr}. 
While intuitive, this generation-based approach introduces inefficiencies and hinders alignment with the goals of ranking. 

\subsection{Generative Reranking with LLMs}\label{sec:related_work-gen}
% \begin{table}[ht]
% \scriptsize
% \centering
% \begin{tabularx}{\linewidth}{lcccc}
% \toprule
% Model & \multicolumn{2}{c}{Bias Mitigation} & {\scriptsize Global} & {Generation } \\
% \cmidrule(lr){2-3}
%       & Positional & Semantic &  Ranking  &   Target                \\
% \midrule
% monoT5     & --       & --       & --           & Token       \\
% RankT5     & --       & --       & --           & Token       \\
% ListT5     & \cmark  & \xmark & \xmark     & Sequence    \\
% RankZephyr & \cmark  & \xmark & \xmark     & Sequence    \\
% PE-Rank    & \xmark & \xmark & \cmark      & Sequence    \\
% FIRST      & \xmark & \xmark & \xmark     & Token       \\
% \midrule
% MVP      & \cmark  & \cmark  & \cmark      & Anchor     \\
% \bottomrule
% \end{tabularx}
% \caption{Comparison of generative reranking models and MVP with respect to bias mitigation, full-ranking capability, and decoding strategy.}
% \label{tab:model_comparison}
% \end{table}

\newcommand{\VCell}[1]{\raisebox{-0.6\height}{#1}}

\begin{table}[t]
\small
\centering
  \setlength{\tabcolsep}{3pt}  % 원래는 6pt 정도, 값이 작을수록 간격이 좁아집니다.

\begin{tabular}{@{}lcccc@{}}
\toprule
\multirow{2}{*}{\VCell{Model}}
  & \multirow{2}{*}{\VCell{%
        \begin{tabular}{c}Global\\Ranking\end{tabular}}}
  & \multirow{2}{*}{\VCell{%
        \begin{tabular}{c}Single Pass\\Decoding\end{tabular}}}
  & \multicolumn{2}{c}{Bias Mitigation}
 \\
\cmidrule(lr){4-5} 
                       &             &               &   Position                                                                        &    Selection                                                                          \\
\midrule
ListT5                 & \xmark & \xmark & \cmark                                                     & \xmark                                                                     \\
RankZephyr             & \xmark & \xmark & \cmark                                                     & \xmark                                                                     \\
PE-Rank                & \cmark & \xmark & \xmark                                                     & \xmark                                                                     \\
FIRST                  & \xmark & \cmark & \xmark                                                     & \xmark                                                                     \\
\midrule
MVP                  & \cmark & \cmark & \cmark                                                     & \cmark     \\                       \bottomrule                                       
\end{tabular}
\caption{Comparison of generative LLM rerankers and MVP with respect to bias mitigation, global ranking capability, and generation target.}
\label{tab:model_comparison}
\end{table}

% To address the limitations discussed in Section~\ref{sec:introduction}, various generative reranking methods have been proposed.
% postion bias 문제를 해결하기 위헤 ListT5~\cite{acl/YoonCKYKH24/ListT5}는 FiD 구조를 활용을 하였으며, RankZephyr~\cite{corr/abs-2312-02724/RankZephyr}는 입력 순서를 섞어주거나, 다양한 문서 수를 학습에 넣어주어 문제를 해결하려고 하였다. 그러나 이 두 논문들은, AutoRegresive 한 방식 및 Global Ranking이 불가능하다는 점에서 inefficient하다.
% PE-Rank~\cite{corr/abs-2406-14848/PE-Rank} compresses each passage into a single token하였으며, Global Ranking이 가능하지만,  but suffers from information loss during compression and projection. 
%  FIRST~\cite{emnlp/ReddyDXSSSJ24/FIRST}는 Single passa decoding, 즉, using the logits of the first generated token 을 활용하여 추론이 가능하여 효율성을 극대화 하였다. global ranking이 불가능하다는 점과 여러 bias에 대한 해결을 하지 못했다. 
%   즉 여러 방면에서 generation-based approach의 문제를 해결하기 위한 연구가 존재하지만, 어떤한 연구도, However, no prior work has simultaneously achieved (i) 여러 Bias를 제거하고, (ii) global ranking이 가능한 형태, (iii) Single pass decoding을 달성하진 못하였다. . To this end, we propose \textbf{MVP}, which unifies all three into a single reranking framework. A comparison with existing methods is summarized in Table~\ref{tab:model_comparison}. 
To address the limitations discussed in Section~\ref{sec:introduction}, various generative reranking methods have been proposed. 
To mitigate the position bias, ListT5~\cite{acl/YoonCKYKH24/ListT5} leverages the FiD architecture, while RankZephyr~\cite{corr/abs-2312-02724/RankZephyr} addresses the issue by shuffling input order and varying the number of passages during training. 
PE-Rank~\cite{corr/abs-2406-14848/PE-Rank} compresses each passage into a single token, allowing global ranking, but suffers from information loss during compression and projection.  
FIRST~\cite{emnlp/ReddyDXSSSJ24/FIRST} improves efficiency via single-pass decoding using the logits of the first generated token, yet supports neither global ranking nor effective bias mitigation. 

While prior work has tackled individual aspects of generation-based reranking, no method has simultaneously achieved (i) mitigation of various biases, (ii) global ranking capability, and (iii) single-pass decoding. A comparison with existing methods is presented in Table~\ref{tab:model_comparison}.

%To this end, we propose \textbf{MVP}, a unified reranking framework that satisfies all three criteria. 

%  takes a listwise perspective, using the logits of the first generated token for both training and inference.

% ListT5~\cite{acl/YoonCKYKH24/ListT5}  employs a Fusion-in-Decoder (FiD) architecture to address postion bias, 
% RankZephyr~\cite{corr/abs-2312-02724/RankZephyr} 는 다양한 문서 수와 query를 사용하는 학습 전략을 통해 postion bias에 대한 문제를 해결하고
% though it remains inefficient due to autoregressive decoding. 

% follow this approach. PE-Rank~\cite{corr/abs-2406-14848/PE-Rank} compresses each passage into a single token, but suffers from information loss during compression and projection. 

% \noindent\textbf{Logit-Based Estimation.} 
% Following models perform reranking based on the logits for generating a single token. MonoT5~\cite{emnlp/NogueiraJPL20/MonoT5} and RankT5~\cite{sigir/Zhuang0J0MLNWB23/RankT5} adopt a pointwise structure, leveraging the logits of relevance tokens to approximate relevance. FIRST~\cite{emnlp/ReddyDXSSSJ24/FIRST} takes a listwise perspective, using the logits of the first generated token for both training and inference. However, no prior work has simultaneously achieved (i) explicit representation of query–passage relevance, (ii) full listwise ranking, and (iii) non-generative scoring. To this end, we propose \textbf{MVP}, which unifies all three into a single reranking framework. A comparison with existing methods is summarized in Table~\ref{tab:model_comparison}. 

% \input{sec-3-method}
\begin{figure*}[t]
\includegraphics[width=1.0\linewidth]{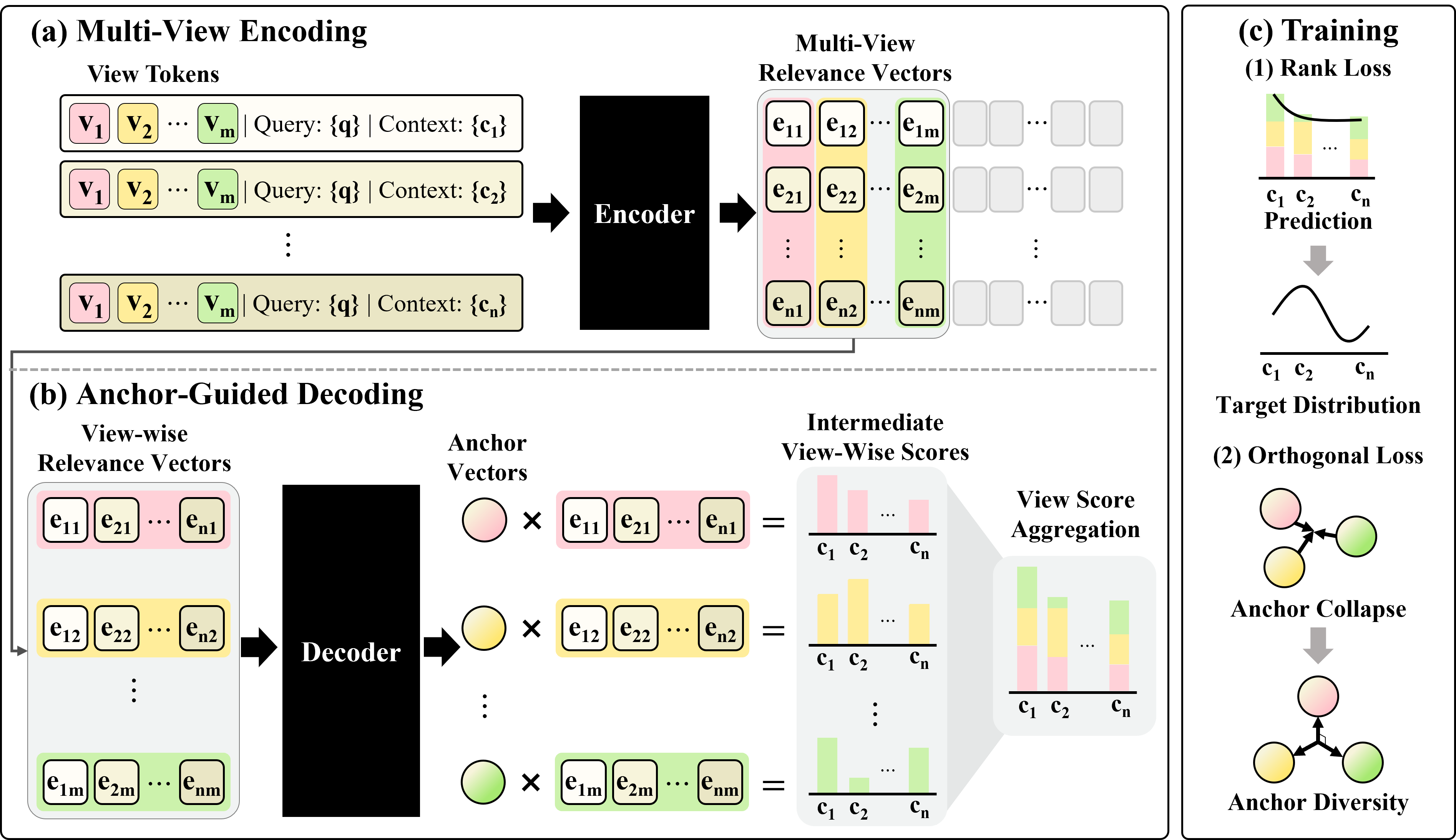}
% \vskip 0.5in
\caption{The overall framework of \textbf{MVP}.
(a) A query-passage pair is encoded into multiple relevance vectors, where each vector represents a distinct view. 
(b) For every view, an anchor vector is generated, and the view-wise relevance score is computed based on its similarity to the corresponding relevance vector. The final score is obtained by aggregating scores across all views.
(c) The model is trained with a ranking loss to match the target distribution and an orthogonality loss to encourage diversity among anchor vectors.
}
\label{fig:fig_MVP_framework}
\vskip -0.1in
\end{figure*}

% \caption{
% The overall framework of \textbf{MVP}. 
% (a) Each query-passage pair is encoded into multiple relevance vectors, where each vector represents a distinct view. 
% (b) For each view, an anchor vector is generated, and relevance score is computed by measuring the similarity between the anchor and its corresponding relevance vector. The final score is obtained by aggregating scores across all views. 
% (c) The model is trained with a ranking loss to match the target distribution and an orthogonality loss to encourage diversity among anchor vectors.
% % }
\section{Proposed Method} \label{sec:method}

% In this section, we present MVP, an efficient reranking framework built on the Fusion-in-Decoder (FiD) architecture. Figure~\ref{fig:fig_MVP_framework} illustrates the overall architecture of \text{MVP}. First, we encode each query–passage pair into multiple special tokens that form query-aware passage embeddings, where each embedding serves as a distinct view, capturing unique relevance information (Section~\ref{sec:obtaining QPE}). These view-wise embeddings are then independently processed by the decoder to produce corresponding anchor vectors, each of which computes a relevance score via a dot product with the passage embedding from the same view (Section~\ref{sec:anchor reranking}). Finally, we train the model solely with a ranking objective, combined with an orthogonality regularization term to ensure that anchor vectors remain distinct (Section~\ref{sec:MVP_training}).

In this section, we propose a novel passage reranking model, \textit{\textbf{M}\textit{ulti-\textbf{V}iew-guided \textbf{P}assage reranking} (\textbf{MVP})}, which is based on the FiD architecture. As shown in Figure~\ref{fig:fig_MVP_framework}, a query-passage pair is encoded into multiple relevance vectors, each capturing a unique relevance signal from a different view (Section~\ref{sec:obtaining QPE}). The decoder generates anchor vectors for each view, which score passages via dot product with their relevance vectors (Section~\ref{sec:anchor reranking}). Finally, we train the model solely with a ranking objective with an orthogonality regularization term to ensure that anchor vectors remain distinct (Section~\ref{sec:MVP_training}).

% View-Specific Encoding with Soft Prompts
% 구조 1. fid (질의-패시지 개별 인코딩)
% 구조 2. Soft prompts (문서 간 토큰 위치는 동일, 하나의 문서 안에서 다른 position embedding은 각 뷰를 의미하게 함)
%-> 결과1. 새로운 특수 토큰으로 초기화된 임베딩은 selection bias에도 강건함.
%-> 결과2. (1) + (2) 시너지 발휘 => position bias를 완전히 제거하여 효과적인 encoding 수행

\subsection{Multi-View Encoding}\label{sec:obtaining QPE}
% 입력값 처리 (spcecial token)

To employ a query-aware passage embedding that summarizes the entire context, we encode each query-passage pair through a set of learnable soft prompts. Given a query $q$ and a set of $n$ candidate passages $[c_1, c_2, \dots, c_n]$, we construct a distinct input prompt $x_i$ by prepending $m$ view tokens $\langle\text{v}_1\rangle, \langle\text{v}_2\rangle, \dots, \langle\text{v}_m\rangle$ to the query and the $i$-th passage. The relative positions of these view tokens are fixed across all passages, ensuring that each $\langle\text{v}_k\rangle$ consistently appears at the same location, regardless of the query and passage content. Meanwhile, each view token in $x_i$ is assigned a unique position embedding, enabling the model to distinguish between views and capture diverse aspects of the query–passage relationship.
\begin{equation}
x_i = \langle \text{v}_1 \rangle \cdots \langle \text{v}_{m} \rangle \mid \text{Query: } q \mid \text{Context: } c_i
\end{equation} \label{eq:add_view}
The FiD encoder processes constructed input $x_i$ to obtain hidden states  $H_i$, where $L$ denotes the length of the input sequence and $d$ denotes the hidden size of the language model. 
\begin{align}
H_i &= \text{FiD}_{\text{encoder}}(x_i), \quad H_i \in \mathbb{R}^{L \times d}
\end{align}\label{eq:FID_encoding}
From these hidden states, we extract the vectors corresponding to each special token $\langle\text{v}_k\rangle$, denoted as  $e_{ik}$, representing distinct views of query-passage relevance:
\begin{align}
e_{ik} &= H_i[\langle \text{v}_k \rangle]\quad\text{for } k = 1, \dots, m
\end{align}\label{eq:hidden_sates}
Consequently, each candidate passage $c_i$ is compressed into a set of $m$ relevance vectors, ${e_{i1}, e_{i2}, \dots, e_{im}}$. The integration of the FiD architecture and position-controlled soft prompts effectively eliminates both position and selection biases, enabling robust and view-specific encoding of query–passage interactions. 

%These multi-view representations are subsequently used for anchor-based scoring (Section~\ref{sec:anchor reranking}).

% (ii) Multi-anchor guided decoding
% Novel reranking strategy(framework)
%   장점 1. single pass decoding, decoder  
%   장점 2. non-generative => 추론과 학습과정에서 lm head를 사용하지 않기 때문에 ranking objective로만 학습 가능
% orthogonal constraints(이름확인)을 적용하여, ranking 학습과정에서 각 view embedding collapse 문제완화
\subsection{Anchor-Guided Decoding} \label{sec:anchor reranking}

To minimize the computational overhead of sequential generation, MVP adopts anchor-guided decoding. Specifically, MVP generates multiple anchor vectors by applying cross-attention over all candidate relevance vectors in the decoder, enabling single-pass inference and global ranking without autoregressive decoding. 
% anchor에 대한 설명 추가 이걸 -> 가능

% 첫문장에 view 마다 계산됨을 추가, generated 표현 대체
% 첫문장 수정

% Each anchor, computed for each view from the encoded query-passage re
% Each anchor, generated from a encoded query-passage representation via a dedicated relevance vector (Section\ref{sec:obtaining QPE}), reflects a distinct relevance perspective. 
Each anchor, derived from the relevance vectors corresponding to each view, represents a distinct perspective of relevance. Specifically, given $n$ candidate passages and their $k$-th view relevance vectors ${e_{1k}, \dots, e_{nk}}$, we construct a matrix $E_k \in \mathbb{R}^{n \times d}$ as the key-value input to the decoder. $E_k$ is then transformed into an anchor vector $a_k$ via cross-attention.
% The decoder then performs cross-attention over $E_k$, yielding a single anchor vector $a_k$:
\vspace*{-0.2em}
\begin{align}
E_k &= [e_{1k}; e_{2k}; \dots; e_{nk}] \in \mathbb{R}^{n \times d} \\
a_k &= \text{FiD}_{\text{decoder}}([\text{BOS}], E_k) \in \mathbb{R}^{1 \times d}
\end{align}
% The relevance score from each view is computed by measuring the similarity between a relevance vector $e_{ik}$ and its corresponding anchor $a_k$, and the final score $s_i$ is obtained by averaging scores across all $m$ views:
The relevance score from each view is computed by measuring similarity between a relevance vector $e_{ik}$ and its anchor $a_k$, and the final score $s_i$ is obtained by averaging across all $m$ views:
\vspace*{-0.2em}
\begin{align}
s_i = \frac{1}{m} \sum_{k=1}^{m} \langle a_k, e_{ik} \rangle
\end{align}

By utilizing multiple anchors, the model effectively evaluates candidate passages from diverse semantic views, enabling efficient and accurate scoring without the need for ranking list generation. Importantly, this direct scoring mechanism removes the need to compute token-level logits, enabling both training and inference to rely solely on relevance-based objectives.
\vspace*{-0.2em}

\subsection{Training} \label{sec:MVP_training}
Training MVP involves optimizing two complementary objectives that jointly enhance ranking accuracy and representational diversity. The first is a ranking objective that enables the model to learn the relevance order of candidate passages (Section~\ref{sec:ranking objective}). The second is an orthogonality objective that encourages each anchor to capture a distinct perspective on relevance (Section~\ref{sec:orthogonal objective}).
\subsection{Ranking Loss} \label{sec:ranking objective}
As the ranking objective to train MVP, we adopt the ListNet loss~\cite{icml/CaoQLTL07/ListNET}, which enables the predicted ranking scores to align with the ground-truth relevance order. Given $n$ candidate passages and their ground-truth ranks $r_i \in [1, 2, \dots, n]$ (with $r_i = 1$ indicating the most relevant), each rank is converted into a relevance score using a reciprocal transformation, \ie, $y_i = 1 / r_i$.
We then apply a temperature-scaled softmax to both ground-truth scores ${y_i}$ and predicted scores ${s_i}$ to obtain probability distributions, where $\tau$ is a temperature hyperparameter:
% \vspace*{-0.1em}
\begin{align}
P(y_i) &= \frac{\exp(y_{i} / \tau)}{\sum_{j=1}^{n} \exp(y_{j} / \tau)} \\
P(s_i) &= \frac{\exp(s_{i} / \tau)}{\sum_{j=1}^{n} \exp(s_{j} / \tau)}
% P(z)_i &= \frac{\exp(z_{i} / \tau)}{\sum_{j=1}^{n} \exp(z_{j} / \tau)}
\end{align}

The listwise ranking objective used to approximate the predicted probability for the $i$-th passage is defined as follows:
\begin{align}
\mathcal{L}_{\text{Rank}} &= - \sum_{i=1}^{n} P(y_i) \log P(s_i)\label{eq:listnet}
% \mathcal{L}_{\text{Rank}} &= - \sum_{i=1}^{n} P(y)_i \log P(s)_i\label{eq:listnet}
\end{align}

\subsubsection{Orthogonal Loss} \label{sec:orthogonal objective}

Since MVP computes the relevance score for each passage by leveraging multiple anchor vectors, each anchor has to capture a distinct and complementary view of the query-passage relationship. To this end, we introduce an orthogonal regularization loss that promotes diversity among anchor vectors, inspired by the Orthogonal Projection Loss (OPL)~\cite{iccv/RanasingheNH0K21/orthogonal}, which encourages orthogonality in feature representations. The loss is defined as:
\begin{align}
\mathcal{L}_{\text{Orthogonal}} &= \sum_{k=1}^{m} \sum_{\substack{l=1 \\ l \ne k}}^{m}  [a_k, a_l]^2 \\
[x_i, x_j] &= \frac{x_i \cdot x_j}{\|x_i\|_2 \cdot \|x_j\|_2}
\end{align}

Here, \([\,\cdot\,,\,\cdot\,]\) denotes the cosine similarity operator, and \(\|\cdot\|_2\) represents the L2 norm.

This regularization encourages anchor vectors to remain in distinct directions, guiding the encoding stage to capture diverse semantic views across tokens. The final training loss combines the primary ranking loss and the orthogonality regularization term:
\begin{equation}
\mathcal{L} = \mathcal{L}_{\text{Rank}} + \mathcal{L}_{\text{Orthogonal}}
\end{equation}

% 여기에서 Special Token의 개수는 4개를 사용했다고 이야기해야함.

\section{Experiments}\label{sec:exp_setup}
% \usepackage{booktabs}
% \usepackage[normalem]{ulem}
% \begin{table*}[t]\small
% \centering

\begin{table*}[t]\small
\centering
\setlength\tabcolsep{3pt}
\begin{tabular}{@{}l|cc|cccccccccc|c@{}}
\toprule
\textbf{Model} & \textbf{DL19} & \textbf{DL20} & \textbf{Covid} & \textbf{NFCorpus} & \textbf{Signal} & \textbf{News} & \textbf{Robust04} & \textbf{SciFact} & \textbf{Touche} & \textbf{DBPedia} & \textbf{BEIR Avg.} \\ \midrule
MonoT5 (220M)                & 71.5          & 67.0          & 78.3           & \underline{35.7}        & 32.0            & 48.0          & 53.4              & 73.1             & 29.6            & 42.8             & 49.1               \\
RankT5 (220M)                & \underline{72.4}    & \underline{68.3}    & 77.7           & 35.1              & 30.8            & 45.4          & \underline{54.3}        & 73.5             & \underline{37.1}      & \underline{43.7}              & 49.7               \\
ListT5 (220M)                & 71.8          & 68.1          & 78.3           & 35.6              & \textbf{33.5}   & \underline{48.5}    & 52.1              & \underline{74.1}       & 33.4            & \underline{43.7}             & \underline{49.9}         \\ \midrule

% \rowcolor{gray!13}
MVP  (220M)                & \textbf{74.3} & \textbf{69.2} & \textbf{80.2}  & \textbf{36.0}     & \underline{32.7}      & \textbf{49.1} & \textbf{55.1}     & \textbf{75.0}    & \textbf{39.1}   & \textbf{43.8}       & \textbf{51.4}      \\ \midrule
MonoT5 (3B)                             & 71.8          & 68.9          & 79.8          & 37.3          & 32.2          & 48.3          & \underline{58.5}    & 76.3          & 32.5          & 44.8          & 51.2          \\
RankT5 (3B)                             & 72.5          & 70.4          & 81.7          & 37.4          & 31.9          & 49.5          & 58.3          & \textbf{77.1} & \textbf{38.8} & 45.0          & 52.5          \\
ListT5 (3B)                             & 71.8          & 69.1          & \textbf{84.7} & \textbf{37.7} & 33.8          & \textbf{53.2} & 57.8          & \underline{77.0}    & 33.6          & 46.2          & \underline{53.0}    \\
FIRST (7B)                              & 72.4          & \textbf{71.1} & 82.4          & 36.3          & \underline{34.0}    & 52.4          & 54.6          & 75.0          & \underline{38.0}    & \underline{46.3}    & 52.6          \\
PE-Rank (7B)                            & 70.8          & 65.4          & 77.8          & 34.8          & 32.0          & 52.3          & 48.7          & 70.2          & 34.2          & 40.6          & 49.0          \\
RankVicuna (7B) & 66.5          & 66.4          & 79.5          & 32.5          & 33.3          & 45.0          & 47.0          & 68.8          & 32.9          & 44.5          & 48.1          \\
RankZephyr (7B)                         & \underline{73.1}    & \underline{70.8}    & \underline{83.2}    & 36.3          & 31.5          & \underline{52.5}    & 54.3          & 74.9          & 32.4          & 44.5          & 51.4          \\
\midrule
MVP (3B)                              & \textbf{73.5} & \textbf{71.1} & 83.1          & \underline{37.6}    & \textbf{34.2} & 51.2          & \textbf{60.5} & 76.4          & 37.2          & \textbf{46.6} & \textbf{53.3} \\
\bottomrule
\end{tabular}
\caption{Results (nDCG@10) of reranking top-100 passages on TREC and BEIR benchmarks.
The initial candidate passages are retrieved using BM25. The best-performing model in each section is highlighted in \textbf{bold}, and the second-best is marked with \underline{underline}.}
\label{tab:maintable}
\end{table*}
% \vskip -0.2in
\setlength{\tabcolsep}{6pt}  % 기본값으로 복원
% % 이 섹션에서는 먼저 제안하는 모델인 MVP의 학습 및 평가 환경을 설명한 후, 세 가지 주요 실험 결과를 다음과 같은 순서로 제시한다. (1) Section~\ref{sec:performance}에서는 다양한 리랭킹 모델들과의 전반적인 성능 및 효율성을 비교하고, (2) Section~\ref{sec:robustness}에서는 다른 리스트와이즈 모델들과 함께 position bias에 대한 강건성을 분석한다. (3) 마지막으로 Section~\ref{sec:ablation}에서는 제안한 모델의 설계 요소들이 성능에 미치는 영향을 확인하기 위한 ablation 실험을 수행하였다. 별도로 명시하지 않은 경우, 모든 실험 결과는 T5-base 기반 모델을 사용하여 측정되었다.
% In this section, we first describe the training and evaluation setup for our proposed model, MVP. We then present three main experimental results: (1) Comparison of performance and efficiency against various re-ranking models; (2) Robustness to initial order in comparison with other listwise reranking models; (3) Ablation studies to assess the effectiveness of key architectural components. All experimental results for MVP are based on the T5-base model, unless explicitly stated as 3B.

In this section, we first describe the training and evaluation setup for MVP. We then present four main results: (i) overall ranking performance, (ii) efficiency against various reranking models, (iii) robustness to external biases, and (iv) ablation studies on key architectural components. All results for MVP are based on the T5-base model unless otherwise specified as 3B.

\subsection{Experimental Setup}

\noindent\textbf{Datasets.}
We evaluated in-domain performance on the TREC-DL19, DL20~\cite{corr/abs-2003-07820/DL19, corr/abs-2102-07662/DL20}, and assessed zero-shot out-of-domain performance on the BEIR~\cite{corr/abs-2104-08663/BEIR} benchmark, which is designed to evaluate the generalization ability of ranking models. Although BEIR comprises eight diverse datasets, we followed prior work~\cite{emnlp/0001YMWRCYR23/RankGPT} and conducted evaluations on eight datasets with relatively fewer queries. We employed BM25 as the first-stage retrieval model and measured reranking performance using Normalized Discounted Cumulative Gain at rank 10 (nDCG@10). Note that while we use five passages per query during training, at inference we rerank all candidate passages using single-pass decoding without any other sorting algorithms.

% 본 연구에서는 학습 데이터로 MS MARCO Passage Ranking 데이터셋을 사용하였다. 이 데이터셋은 532,761개의 질의(Query)와 약 880만 개의 단락(Passage)으로 구성되어 있으며, 질의와 단락의 관련성은 이진 값(0 또는 1)으로 표시된다. 여기서우리는 RankDistill 논문에서 제공한 파생 데이터를 활용하였다. 이 데이터는 10,000개의 질의에 대해 ColBERTv2(Ref) 검색기를 사용하여 상위 100개의 후보 문서를 추출하고, 이 후보군에 대해 RankZephyr-7B 모델을 통해 정제된 순위 라벨을 부여한 데이터이다. 우리는 해당 데이터에서 각 질의마다 5개의 후보 문서를 비복원 추출로 구성하고, 이를 복원 추출로 100회 반복하여 학습 데이터를 구성하였다. 우리는 학습에서는 하나의 Instance에 5개의 Passage를 사용하지만, 추론 시에는 First retrieval이 제공하는 모든 candidate Passage를 한번에 재순위화 한다. 

% \vspace{1mm}
\noindent
\textbf{Implementation Detail}. To train MVP, we utilized the Rank-DistiLLM~\cite{ecir/SchlattFSZKZSPH25a/RDLLM} dataset, which is constructed from the MS MARCO passage ranking dataset~\cite{/nips/NguyenRSGTMD16/MSMARCO} using 10,000 queries. For each query, the top 100 candidate passages were first retrieved using ColBERTv2~\cite{naacl/SanthanamKSPZ22/colbertv2}, and then reranking these passages with the RankZephyr~\cite{corr/abs-2312-02724/RankZephyr}. To construct a more diverse training set, we sampled 5 candidate passages 100 times per query, resulting in approximately 1 million instances.

We adopted T5-base and T5-3B~\cite{jmlr/RaffelSRLNMZLL20/T5} as our backbone models. For optimization, we applied DeepSpeed Stage 2. For T5-base, we used a batch size of 16, gradient accumulation steps set to 2, a learning rate of 1e-4, and a linear scheduler with a warm-up ratio of 5\%. For T5-3B, we used a batch size of 2, gradient accumulation steps of 16, and a learning rate of 1e-5. The maximum input sequence length was fixed at 256 tokens for both models. Training was conducted for a single epoch, taking approximately 5 hours on 2 × NVIDIA RTX 3090 GPUs for T5-base, and 40 hours on 2 × NVIDIA A6000 GPUs for T5-3B.  
We use $m=4$ special tokens to represent the relevance views, implemented using the T5 tokenizer’s predefined tokens \texttt{<extra\_id\_0>} to \texttt{<extra\_id\_3>},  and set $\tau = 0.8$ to control the sharpness in the ListNet loss. For validation, we use the TREC-DL21 dataset~\cite{trec/Craswell0YCL21DL21/DL21} with nDCG@10 as the validation metric.

% 성능 평가는 TREC-DL19와 DL20을 통해 in-domain 성능을 측정하고, BEIR benchmark에서 zero-shot out-of-domain 성능을 평가하였다. BEIR benchmark는 총 18개의 다양한 도메인으로 구성되지만, 기존 연구(RankGPT 등)를 따라 비교적 적은 질의를 가진 8개 데이터셋에 대해 평가를 진행하였다. 첫 단계 검색 모델로는 BM25를 사용했으며, 재순위화 성능은 Normalized Discounted Cumulative Gain at rank 10 (nDCG@10)을 사용하여 평가하였다. 모든 점수는 소수점 첫째 자리까지 반올림하여 제시하였다.

% BEIR contains
%  18 datasets from different fields with different query requirements,
%  aiming to evaluate the generalization ability of ranking models. Fol
% lowing previous work [33], we conduct evaluations on 8 datasets
%  that contain arelatively small numberofqueries.

\subsection{Ranking Performance} \label{sec:performance}

% \textbf{Performance.} 
% 본 연구에서는 제안하는 모델 MVP의 성능을 기존의 Pointwise 방식(MonoT5\cite{emnlp/NogueiraJPL20/MonoT5}, RankT5\cite{sigir/Zhuang0J0MLNWB23/RankT5}), Listwise 방식(ListT5\cite{acl/YoonCKYKH24/ListT5}) 기반의 리랭킹 모델들과 7B 규모의 대형 언어모델(RankVicuna, RankZephyr, FIRST, PE-Rank)과 비교하였다. 비교 결과는 표 X에 요약되어 있다.
We compare MVP against seven reranking models built on the T5 architecture. Specifically, pointwise models are \textbf{MonoT5}~\cite{emnlp/NogueiraJPL20/MonoT5} and  \textbf{RankT5}~\cite{sigir/Zhuang0J0MLNWB23/RankT5}.  The listwise model is \textbf{ListT5}~\cite{acl/YoonCKYKH24/ListT5}. For 7B-scale rerankers, we employ \textbf{RankVicuna}~\cite{corr/abs-2309-15088/RankVicuna}, \textbf{RankZephyr}~\cite{corr/abs-2312-02724/RankZephyr}, \textbf{FIRST}~\cite{emnlp/ReddyDXSSSJ24/FIRST}, and \textbf{PE-Rank}~\cite{corr/abs-2406-14848/PE-Rank}. Note that MVP is based on 220M and 3B base models.

Table~\ref{tab:maintable} reports the overall result. When evaluated at the T5-base scale, MVP outperforms other baselines of the same model size across most datasets. On the BEIR benchmark, MVP achieves an average nDCG@10 score of 51.4, surpassing MonoT5, RankT5, and ListT5 by 2.3, 1.7, and 1.5 points. This performance is also comparable to that of RankZephyr (7B), a much larger model. On the TREC-DL19 and DL20 datasets, MVP also exceeds RankT5 by 1.9 and 0.9 points, respectively.

We also compare the 3B variant of MVP with large-scale (3B-7B) LLM-based reranking models. MVP-3B achieves nDCG@10 scores of 73.5 on DL19, 71.1 on DL20, and 53.3 on the BEIR average, outperforming all other models at the 3B and 7B scales. These results suggest that the architectural advantages of MVP generalize well to larger model configurations.
%, supporting its scalability and competitive performance.

\subsection{Efficiency}

\begin{figure}[t]
\includegraphics[width=0.95\linewidth]{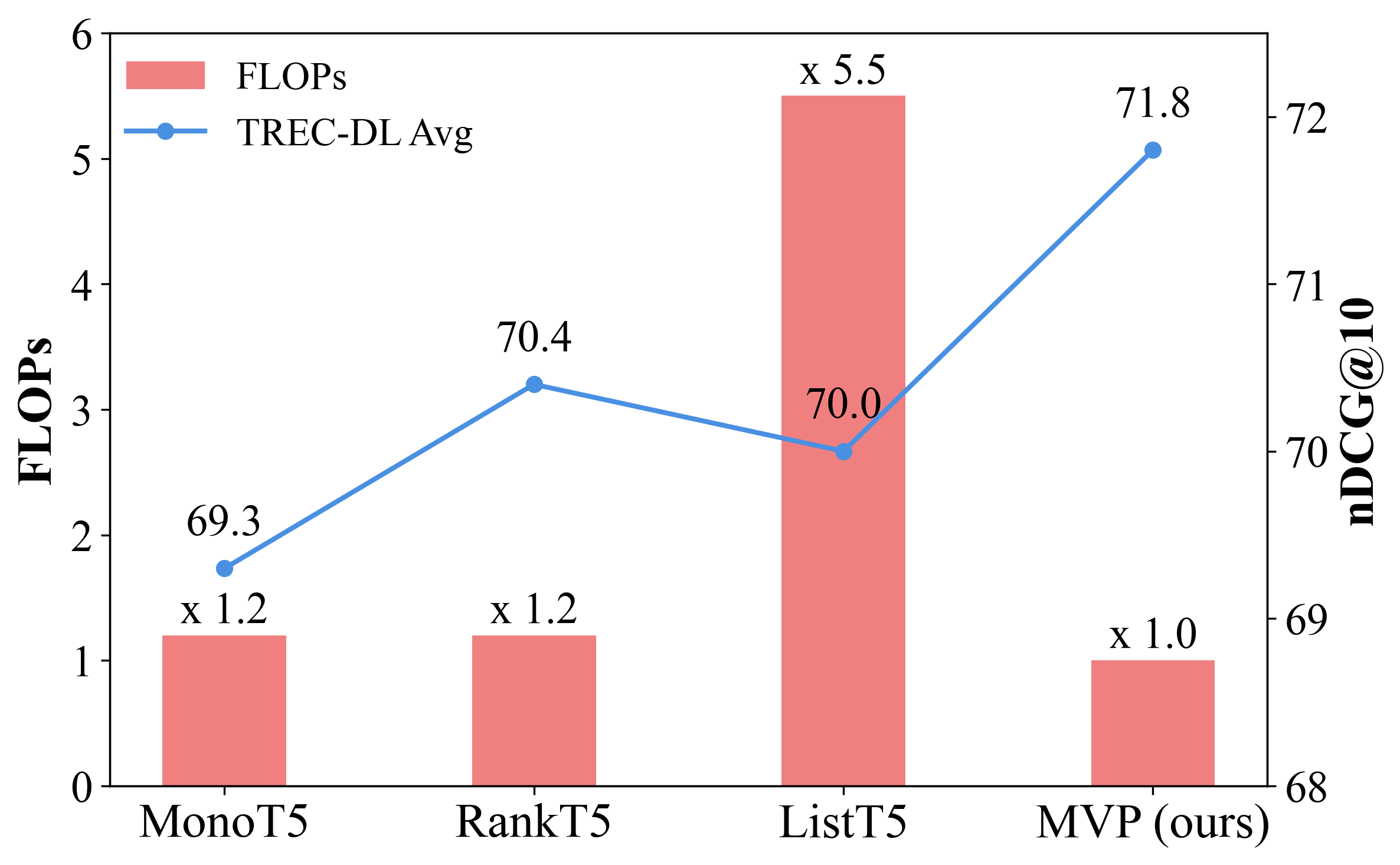}
\caption{Real-time FLOPs comparison of the models. The reported performance is averaged over DL19 and DL20.}
\vskip -0.2in
\label{fig:flops}
\end{figure}
% ~\cite{eacl/IzacardG21/FiD}

%  Real-time FLOPs comparison of the models
% on T5-base, including DuoT5 and the sliding window
% variants of LISTT5. The reported BEIR performance is
% averaged from a subset of BEIR, same as in Tab. 3.
% \textbf{Flops} 모델의 효율성을 정량적으로 평가하기 위해, DeepSpeed의 FlopsProfiler를 활용하여 각 모델의 FLOPs(Floating Point Operations)를 측정하였다. 실험은 TREC-DL19 데이터셋의 43개 질의를 대상으로 수행되었으며, 각 질의에 대해 BM25로 검색된 상위 100개의 후보 문서 중 Top-10을 재순위화하는 과정에서의 연산량을 측정하였다.입력 시퀀스의 길이는 256 토큰으로 고정하였고, 모든 모델은 T5-base를 백본으로 사용하였다. 비교의 기준으로 MVP의 FLOPs를 1.0으로 정규화한 후, 다른 모델들의 상대적인 FLOPs를 산출하였다. 성능 평가는 TREC-DL19 및 DL20에서의 nDCG@10 점수를 기준으로 진행되었다.

% MVP의 강점은, query-passage를 multiple relevance vector로 표현하고, anchor guided decoding을 통해, 높은 성능과 effeiency를 크게 개선한 것이다.
% The key strength of MVP lies in its ability to efficiently compress query-passage into compact representations and leverage multi-anchor guided decoding to achieve strong performance while significantly reducing computational cost.  
The key strength of MVP lies in its ability to represent each query-passage pair with multiple relevance vectors and to perform anchor-guided decoding, achieving both high effectiveness and significantly improved efficiency. To empirically validate efficiency, we report both floating-point operations (FLOPs) and latency. All experiments are conducted on a 24GB NVIDIA RTX 3090 GPU.

\noindent\textbf{FLOPs.} To assess the computational efficiency of each model, we measured FLOPs using DeepSpeed's FLOPs Profiler\footnote{We use DeepSpeed’s FLOPs profiler for measurement: \url{https://github.com/microsoft/DeepSpeed}}. The evaluation was conducted on 43 queries from the DL19 dataset. Following the prior work~\cite{acl/YoonCKYKH24/ListT5}, we measured the FLOPs required to determine the top 10 passages out of BM25-Top100 candidates. The input sequence length was set to 256 tokens. For ease of comparison,  we normalized MVP’s FLOPs to 1.0\footnote{The exact FLOPs value is 197,983,445,625,792.} with the relative FLOPs of other models computed accordingly. 

As illustrated in Figure~\ref{fig:flops}, MVP achieves the lowest computational cost among all models while outperforming them in ranking quality. Compared to ListT5, it reduces FLOPs by approximately 82\%. Notably, MVP also consumes fewer operations than pointwise models such as MonoT5 and RankT5, despite delivering stronger reranking performance.
% This computational efficiency results from MVP’s design strategy, which compresses the query-passage into $m$ special tokens and employs an anchor-based scoring strategies to evaluate all candidates simultaneously, effectively eliminating the need for repeated generation.

% \textbf{Latency} 모델의 실사용 가능성을 평가하기 위해, 각 모델에 대해 Top-10 후보 문서 재순위화 시 소요되는 추론 시간(Latency)을 측정하였다. Latency는 전체 소요 시간을 질의 수로 나눈 평균값으로 정의되며, 단위는 초(Seconds)이다. 실험은 TREC-DL19, DL20, TREC-COVID, NEWS 데이터셋을 대상으로 동일한 조건에서 수행되었으며, 하드웨어는 NVIDIA RTX 3090 GPU를 사용하였다. 공정한 비교를 위해, 일부 모델이 지원하는 vLLM 기반의 추론 가속 기능은 모두 비활성화하였으며, 모든 측정은 모델의 순수한 추론 속도만을 반영하였다. 측정 결과를 Figure~\ref{fig:latency}에 존재한다.
\begin{figure*}[t]
\centering
\includegraphics[width=1.0\textwidth]{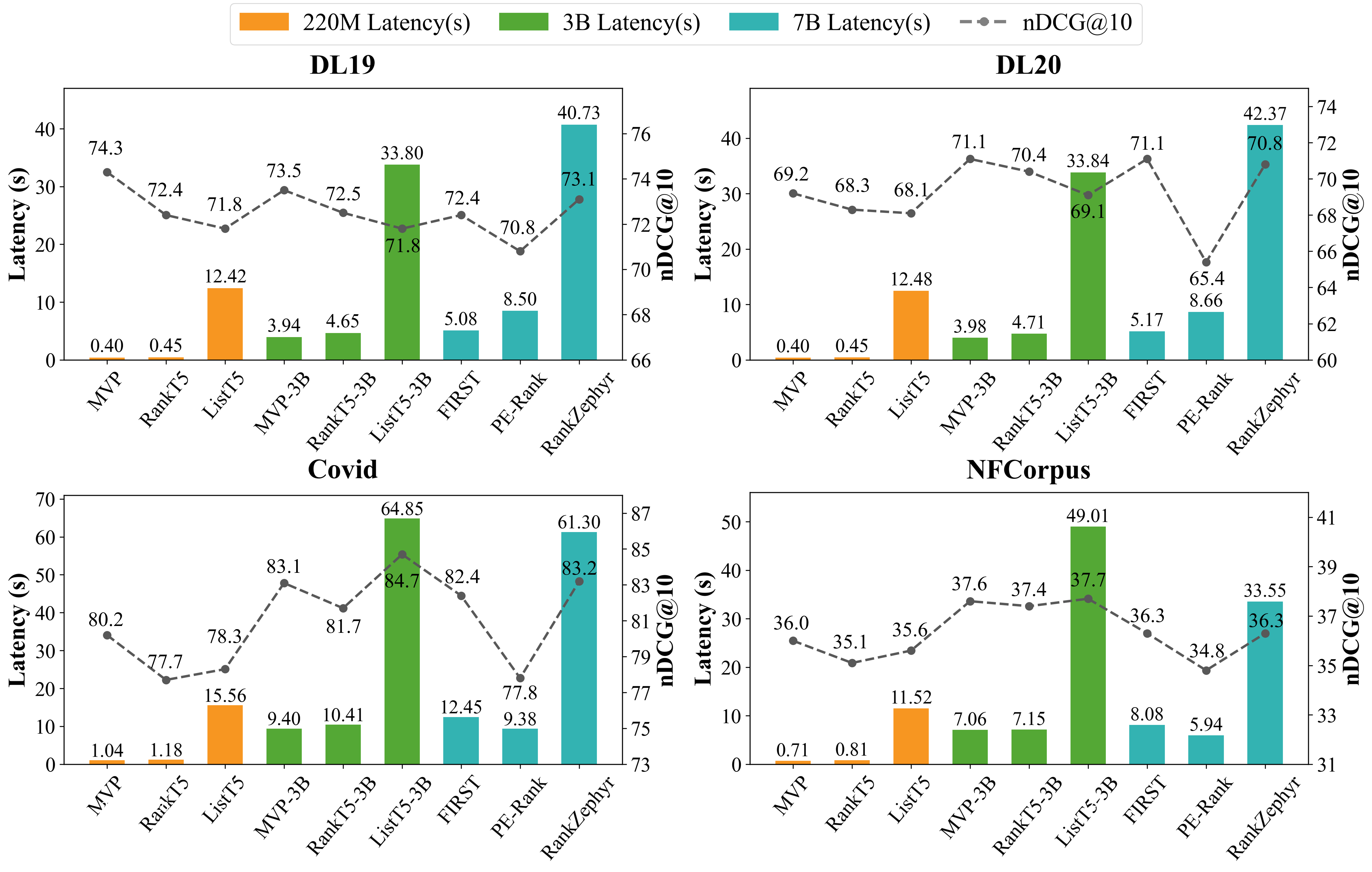}
\caption{
Ranking performance (nDCG@10) for the reranker’s latency (s).  Latency indicates the average time required to rerank a single query.
}
\label{fig:latency}
\vskip -0.1in
\end{figure*}

\noindent\textbf{Latency.}
We also measure the latency required to determine the top-10 passages from BM25-Top100 candidate passages. Latency is defined as the average time per query, measured in seconds. Our experiments are conducted on DL19 and DL20, along with two datasets from the BEIR benchmark: Covid and NFCorpus. For fair comparison, all vLLM acceleration features are disabled, ensuring that the latency reflects the raw inference time of each model. 

The results in Figure~\ref{fig:latency} show that MVP achieves faster inference than existing listwise models across all datasets, even surpassing the pointwise model RankT5. Specifically on DL19, it achieves 100× faster than RankZephyr and 12.7× faster than FIRST, while maintaining comparable ranking performance. At the larger scale (3B-7B), MVP-3B offers a favorable trade-off between latency and ranking accuracy. Notably, even compared to FIRST, a well-balanced 7B model, MVP-3B achieves faster inference and better accuracy.

In summary, the FLOPs and latency results confirm that MVP is both efficient and effective for real-time reranking. The scoring strategy of MVP enables simultaneous evaluation of all candidates without repeated decoding, eliminating redundancy and supporting strong ranking performance.

\begingroup                      % ← 표 안으로 폰트/간격 범위 한정
\setlength{\tabcolsep}{3.6pt}
\renewcommand{\arraystretch}{0.95}

\begin{table}[t]
\centering
\footnotesize                    % ← 표 내부에서만 폰트 축소
\begin{adjustbox}{width=\columnwidth} % ← 한 칼럼 폭에 맞게 자동 축소
\begin{tabular}{l|l|cccc}
\toprule
\textbf{Model} & \textbf{Order} & \textbf{DL19} & \textbf{DL20} & \textbf{News} & \textbf{Average} \\
\midrule
\multicolumn{6}{l}{\textbf{Candidate Permutations}} \\
\midrule
\multirow{3}{*}{MVP}
    & Orig. & 74.3 & 69.2 & 49.1 & 64.2 \\
    & Shuf. & 74.3 & 69.2 & 49.1 & 64.2 {\scriptsize (±0.0)} \\
    & Rev. & 74.3 & 69.2 & 49.1 & 64.2 {\scriptsize (±0.0)} \\
\midrule
\multirow{3}{*}{\shortstack[l]{RankZephyr\\{\scriptsize\itshape(sw: w=20, s=10)}}}
    & Orig. & 73.1 & 70.8 & 52.5 & 65.5 \\
    & Shuf. & 73.1 & 70.7 & 51.3 & 65.0 {\scriptsize (–0.4)} \\
    & Rev.  & 72.1 & 71.5 & 51.8 & 65.1 {\scriptsize (–0.3)} \\
\midrule
\multirow{3}{*}{\shortstack[l]{FIRST\\{\scriptsize\itshape(sw: w=20, s=10)}}}
    & Orig. & 72.4 & 71.1 & 52.4 & 65.3 \\
    & Shuf. & 70.0 & 69.4 & 47.3 & 62.2 {\scriptsize (–3.1)} \\
    & Rev.  & 67.5 & 68.3 & 42.4 & 59.4 {\scriptsize (–5.9)} \\
\midrule
\multicolumn{6}{l}{\textbf{Identifier Permutations}} % ← 오타 수정
\\ \midrule
\multirow{1}{*}{MVP}
    & \multicolumn{1}{c|}{--} & 74.3 & 69.2 & 49.1 & 64.2 \\
\midrule
\multirow{3}{*}{\shortstack[l]{RankZephyr\\{\scriptsize\itshape(sw: w=20, s=10)}}}
    & Orig. & 73.1 & 70.8 & 52.5 & 65.5 \\
    & Shuf. & 71.3 & 67.3 & 46.7 & 61.8 {\scriptsize (–3.7)} \\
    & Rev.  & 69.3 & 63.9 & 47.2 & 60.1 {\scriptsize (–5.3)} \\
\midrule
\multirow{3}{*}{\shortstack[l]{FIRST\\{\scriptsize\itshape(sw: w=20, s=10)}}}
    & Orig. & 72.4 & 71.1 & 52.4 & 65.3 \\
    & Shuf. & 71.2 & 69.2 & 49.1 & 63.2 {\scriptsize (–2.1)} \\
    & Rev.  & 71.0 & 68.2 & 48.5 & 62.6 {\scriptsize (–2.7)} \\
\bottomrule
\end{tabular}
\end{adjustbox}
\caption{nDCG@10 across candidate and identifier permutations. Values in parentheses indicate the change relative to the Original order. \textit{sw} denotes the sliding-window setting, with window size ($w$) and stride ($s$) following prior work. Results for the Shuffle setting are averaged over three random seeds.}
\label{tab:passage_order}
\end{table}
\endgroup

\subsection{Robustness to External Biases} \label{sec:robustness}
Most listwise rerankers are sensitive to prompt design—specifically the initial passage order and the choice of passage identifiers—leading to position and selection biases. We evaluate whether our model eliminates these effects under various listwise prompts on DL19, DL20, and News. A detailed analysis is provided in Appendix~\ref{sec:bias analysis}

\vspace{0.5em}
\noindent\textbf{Position Bias.}
% The results shows that MVP exhibits strong robustness under various candidate permutations, indicating effecive mitigation of position bias. 이것은 각 query-passage pair가 individual한 prompt로 구성되고, seperately하게 encoded 되기 때문에, 각 view token들이 shared position embeddings을 사용하기 때문이다.
% \subsubsection{Position Bias}
To evaluate position bias, we manipulate the initial passage order of the BM25 top-100 candidates while keeping identifier tokens fixed within a single reranking window. We consider three configurations: Orig., the BM25 relevance order; Rev., the reversed order; and Shuf., a random permutation. The results are shown in the upper part of Table~\ref{tab:passage_order}.

The results indicate that MVP is robust under different candidate permutations, effectively mitigating position bias. This robustness results from our design choice: each query–passage pair is constructed as an individual prompt and encoded separately, resulting in shared position embeddings of view tokens across all passages.

\vspace{0.5em}
% \subsubsection{Selection Bias}
\noindent\textbf{Selection Bias.}
For selection bias, we fix the candidate order to the BM25 relevance order and manipulate the assignment of identifier tokens. We use configurations similar to the position bias experiment: Orig., the original assignment; Rev., a reversed identifier assignment (e.g.,"[id\textsubscript{20}]: context\textsubscript{1}, [id\textsubscript{19}]: context\textsubscript{2}, \dots, [id\textsubscript{1}]: context\textsubscript{20}"); and Shuf., a random permutation. The results are shown in the lower part of Table~\ref{tab:passage_order}.

Unlike other baselines, MVP does not rely on natural language identifiers. Instead, all query–passage pairs share the same view tokens, rendering identifier permutations inapplicable and effectively eliminating selection bias.

\subsection{Ablation Study}\label{sec:ablation}
To evaluate the impact of key architectural components on model performance, we design several model variants and perform ablation studies. 

\subsubsection{Training Strategies}\label{sec:training strategies}
To investigate the impact of each component in MVP, we perform ablation experiments by removing two key design elements: (i) orthogonality regularization among anchors and (ii) the use of multi-view encoding. The results are reported in Table~\ref{tab:ablation_training}.
% \begin{table}[t]\small
% \centering
% % \begin{adjustbox}{width=1\columnwidth,center}
% \begin{tabular}{c|c|c}
% \toprule
% Variants                   & NQ dev & \# tokens \\ \midrule
% R2C                        & \textbf{58.44}   & 483       \\ \midrule
% R2C w/ chunk only          & 57.01            & 485       \\
% R2C w/ sentence only       & 57.58            & 480       \\
% R2C w/ tokens only         & 51.86            & 478       \\ \midrule
% T5-base initialize         & 44.88            & 483       \\
% w/ decoder last layer only & 56.73            & 483       \\
% unit aggregation: max      & 58.39            & 483       \\
% unit aggregation: sum      & 57.58            & 482       \\ \bottomrule
% \end{tabular}
% % \end{adjustbox}
% \caption{Ablation study of R2C on the NQ dev dataset. The results are based on LLaMA2-7B with 6x compressed prompts.}
% \label{tab:ablation}
% \end{table}

\begin{table}[t]
\centering
\small
\begin{tabular}{lccc}
\toprule
\textbf{Model} & \textbf{DL19} & \textbf{DL20} & \textbf{BEIR Avg.} \\
\midrule
MVP & \textbf{74.3} & \textbf{69.2} & \textbf{51.4} \\
w/o $\mathcal{L}_{\text{Orthogonal}}$ & 73.6 & 66.7 & 50.7 \\
w/o Multi-view Encoding & 73.8 & 68.8 & 50.9 \\
\bottomrule
\end{tabular}
\caption{nDCG@10 for MVP and its ablations on different training strategies. See Table~\ref{tab:ablation_full_beir} for full results.}
\label{tab:ablation_training}
%\vskip -0.1in
\end{table}

% (1) : Anchor 간의 정규화 항을 제거한 경우, 전반적인 성능 저하가 확인되었다. 이는 각 View가 동일한 공간에서 유사하게 작동하면서 상호보완보다는 중복 강조되는 정보로 학습되기 때문으로 해석된다.
\noindent\textbf{w/o Orthogonality.}  
Removing the orthogonality regularization among anchor vectors consistently degrades performance across datasets. This suggests that, in the absence of this constraint, different anchors tend to collapse into similar directions within the embedding space, leading to redundant rather than complementary representations. A detailed analysis of anchor vector similarities is provided in Appendix~\ref{sec:view similiarity}.

% 단일 Token 사용: Anchor를 하나의 Special Token으로 제한한 경우, 전 데이터셋에서 약 0.4~0.5 포인트의 성능 감소가 관측되었다. 이는 Query와 Passage 정보를 하나의 Token에 압축하면서 발생한 정보 손실에 기인한 것으로 판단된다.
\noindent\textbf{w/o Multiple Token.}
Using a single special token to represent relevance results in a 0.4--0.5 point drop in performance on average. This degradation is attributed to the limited capacity of a single token to capture both query and passage information, leading to a loss of discriminative features. 
% This degradation is attributed to the representational bottleneck caused by compressing both query and passage information into a single token, which likely leads to loss of discriminative features.

\begin{figure}[t]
\includegraphics[width=1.0\linewidth]{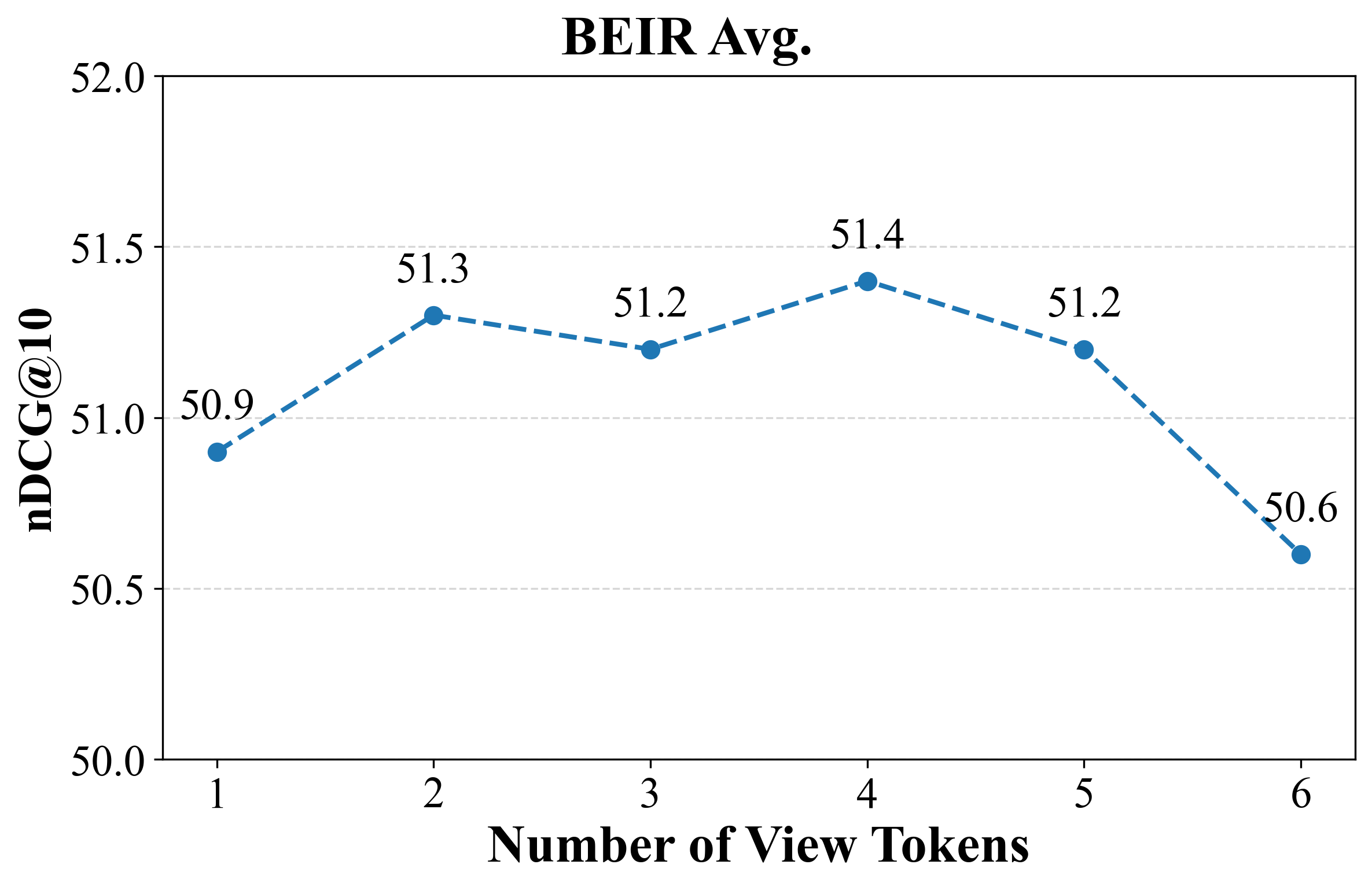}
\caption{Average nDCG@10 on BEIR with respect to the number of view tokens.}
\label{fig:fig_MVP_num}
% \vskip -0.1in
\end{figure}
\subsubsection{Number of View Tokens}
To analyze the impact of the number of view tokens on model performance, we varied the number of relevance tokens from 1 to 6 and evaluated the average performance across BEIR datasets. As illustrated in Figure~\ref{fig:fig_MVP_num}, incorporating multiple views leads to improved performance up to a certain point, beyond which performance begins to degrade.
This suggests that, while orthogonality regularization encourages representational diversity, an excessive number of view tokens may introduce less informative signals, degrading ranking performance.
% View(Anchor)의 수가 모델 성능에 미치는 영향을 분석하기 위해, Relevance Token의 개수를 1개부터 6개까지 변화시키며 실험을 수행하였다. 실험은 8개의 BEIR 데이터를 활용하여 진행을 하였다. As illustrated in Figure~\ref{fig:fig_MVP_num}, 4개일 때 가장 높은 성능을 보였다. 경향을 보면 Multi-view를 활용하는 경우 성능이 상승하지만, 너무 많은 View를 사용하는 경우 오히려 성능이 하락을 한다. It suggests that while orthogonality regularization encourages representational diversity across views, introducing too many view tokens potentially increase of incorporating less informative signals, which can degrade overall ranking performance.
% Figure~\ref{fig:fig_MVP_num}에서 확인할 수 있듯이, TREC 및 BEIR 평균 성능 모두 4개의 Token을 사용할 때 가장 우수한 결과를 보였다. 한편, Token 수가 지나치게 많아질 경우 오히려 성능이 하락하는 경향이 나타났는데, 이는 Orthogonality 정규화를 통해 View 간의 다양성을 유도하게 되는데, 과도한 View 수가 노이즈성 정보까지 포함하게 되어, 최종 관련도 판단에 부정적인 영향을 줄 수 있기 때문으로 해석된다.

% We also examine how the number of view tokens affects model performance by varying it from 1 to 6. As illustrated in Figure~\ref{fig:fig_MVP_num}, the model achieves best performance when using 4 view tokens. However, increasing the number of tokens beyond this point leads to a performance decline. It suggests that while orthogonality regularization encourages representational diversity across views, introducing too many view tokens potentially increase of incorporating less informative signals, which can degrade overall ranking performance.
% \input{sec-results}
\section{Conclusion}\label{sec:conclusion}

We presented \textbf{MVP}, a novel passage reranking model that addresses key limitations of listwise LLM-based approaches, including high computational cost and sensitivity to external biases. By leveraging multi-view encoding through soft prompts and anchor-guided decoding, MVP captures diverse relevance signals efficiently via compact context embeddings, enabling all candidate passages to be evaluated in a single pass, making it particularly well-suited for real-world retrieval scenarios.
Experimental results show that MVP, with only 220M parameters, matches or surpasses the performance of 7B-scale models while reducing inference latency up to 100×. Moreover, its 3B variant achieves state-of-the-art results on both in-domain and out-of-domain benchmarks.

% 이 연구에서 우리는 multi-view query passae encoding과 multi-anchor guided decoding을 활용하는 효율적이면서도 효과적인 novel listwise reranking model인 MVP를 제안한다. 실험 결과 MVP는 220M parameter로도 7B scale 모델들에 비견되는 performance와 더불어 Inference latency에서 12.7 x reduction을 이루어냈으며, 3B 크기로는 in domain 및 out-domain benchmark에서 stae-of-art 결과를 보였다. 또한 Initial ordering에서도 전혀 영향을 받지 않는 모습을 보였다. 
\section{Limitations}\label{sec:limitation}
While MVP employs a fixed number of views across all datasets—a simple and generally effective strategy—using fewer views in some cases can reduce redundancy and improve performance. In addition, MVP aggregates relevance scores by assigning equal weights to all views. Although this uniform aggregation is straightforward, it may overlook the fact that different queries can benefit more from certain views than others. Exploring dynamic view selection or learning query-specific view weights remains a promising direction for future work. 

% f

% 1. MVP는 모든 데이터셋마다 동일한 수의 view를 사용함. 가장 편한 방법이며 일반적으로 가장 높은 성능을 보이지만 일부 데이터셋에서는 그 수가 줄어드는 경우 성능이 더 좋은 경우도 존재한다. 
% 2. MVP는 평가를 할 때 모든 View를 동일한 가중치를 두고 평가를 진행함. 이는 aggregation을 하기에 가장 직관적인 방법이지만, 쿼리마다 중요하게 보는 view가 다를 수도 있음. 만약 쿼리마다 중요한 view를 선택하고, 해당 view의 가중치를 더 높게 준다면 성능이 개선될 수 있으며, 이것은 Future work로 넘긴다.

\section*{Ethics Statement}

This work fully respects ACL's ethical guidelines. We have utilized scientific resources available for research under liberal licenses, and our use of these tools is consistent with their intended applications.

\section*{Acknowledgments} 
% 중견 / 인공지능학과 / 나비효과
This work was supported by the Institute of Information \& communications Technology Planning \& Evaluation (IITP) grant and the National Research Foundation of Korea (NRF) grant funded by the Korea government (MSIT) (No. NRF-RS-2025-00564083, IITP-RS-2019-II190421, IITP-RS-2022-II220680).

% \bibliographystyle{acl_natbib}
\bibliography{references}

\begin{thebibliography}{34}
\expandafter\ifx\csname natexlab\endcsname\relax\def\natexlab#1{#1}\fi

\bibitem[{Cao et~al.(2007)Cao, Qin, Liu, Tsai, and Li}]{icml/CaoQLTL07/ListNET}
Zhe Cao, Tao Qin, Tie{-}Yan Liu, Ming{-}Feng Tsai, and Hang Li. 2007.
\newblock \href {https://doi.org/10.1145/1273496.1273513} {Learning to rank: from pairwise approach to listwise approach}.
\newblock In \emph{Machine Learning, Proceedings of the Twenty-Fourth International Conference {(ICML} 2007), Corvallis, Oregon, USA, June 20-24, 2007}, pages 129--136.

\bibitem[{Cho et~al.(2023)Cho, Jeong, Seo, and Park}]{acl/ChoJSP23/Co-Prompt}
Sukmin Cho, Soyeong Jeong, Jeongyeon Seo, and Jong~C. Park. 2023.
\newblock \href {https://doi.org/10.18653/v1/2023.findings-acl.61} {Discrete prompt optimization via constrained generation for zero-shot re-ranker}.
\newblock In \emph{Findings of the Association for Computational Linguistics: {ACL} 2023, Toronto, Canada, July 9-14, 2023}, pages 960--971. Association for Computational Linguistics.

\bibitem[{Craswell et~al.(2021{\natexlab{a}})Craswell, Mitra, Yilmaz, and Campos}]{corr/abs-2102-07662/DL20}
Nick Craswell, Bhaskar Mitra, Emine Yilmaz, and Daniel Campos. 2021{\natexlab{a}}.
\newblock \href {https://arxiv.org/abs/2102.07662} {Overview of the {TREC} 2020 deep learning track}.
\newblock \emph{CoRR}, abs/2102.07662.

\bibitem[{Craswell et~al.(2021{\natexlab{b}})Craswell, Mitra, Yilmaz, Campos, and Lin}]{trec/Craswell0YCL21DL21/DL21}
Nick Craswell, Bhaskar Mitra, Emine Yilmaz, Daniel Campos, and Jimmy Lin. 2021{\natexlab{b}}.
\newblock \href {https://trec.nist.gov/pubs/trec30/papers/Overview-DL.pdf} {Overview of the {TREC} 2021 deep learning track}.
\newblock In \emph{Proceedings of the Thirtieth Text REtrieval Conference, {TREC} 2021, online, November 15-19, 2021}, volume 500-335 of \emph{{NIST} Special Publication}. National Institute of Standards and Technology {(NIST)}.

\bibitem[{Craswell et~al.(2020)Craswell, Mitra, Yilmaz, Campos, and Voorhees}]{corr/abs-2003-07820/DL19}
Nick Craswell, Bhaskar Mitra, Emine Yilmaz, Daniel Campos, and Ellen~M. Voorhees. 2020.
\newblock \href {https://arxiv.org/abs/2003.07820} {Overview of the {TREC} 2019 deep learning track}.
\newblock \emph{CoRR}, abs/2003.07820.

\bibitem[{Dai et~al.(2024)Dai, Xu, Xu, Pang, Dong, and Xu}]{kdd/DaiXXPDX24/BIAS}
Sunhao Dai, Chen Xu, Shicheng Xu, Liang Pang, Zhenhua Dong, and Jun Xu. 2024.
\newblock \href {https://doi.org/10.1145/3637528.3671458} {Bias and unfairness in information retrieval systems: New challenges in the {LLM} era}.
\newblock In \emph{Proceedings of the 30th {ACM} {SIGKDD} Conference on Knowledge Discovery and Data Mining, {KDD} 2024, Barcelona, Spain, August 25-29, 2024}, pages 6437--6447. {ACM}.

\bibitem[{Hofst{\"{a}}tter et~al.(2023)Hofst{\"{a}}tter, Chen, Raman, and Zamani}]{sigir/HofstatterC0Z23/fid-light}
Sebastian Hofst{\"{a}}tter, Jiecao Chen, Karthik Raman, and Hamed Zamani. 2023.
\newblock \href {https://doi.org/10.1145/3539618.3591687} {Fid-light: Efficient and effective retrieval-augmented text generation}.
\newblock In \emph{Proceedings of the 46th International {ACM} {SIGIR} Conference on Research and Development in Information Retrieval, {SIGIR} 2023, Taipei, Taiwan, July 23-27, 2023}, pages 1437--1447. {ACM}.

\bibitem[{Izacard and Grave(2020)}]{corr/abs-2007-01282/FiD}
Gautier Izacard and Edouard Grave. 2020.
\newblock \href {https://arxiv.org/abs/2007.01282} {Leveraging passage retrieval with generative models for open domain question answering}.
\newblock \emph{CoRR}.

\bibitem[{Karpukhin et~al.(2020)Karpukhin, Oguz, Min, Lewis, Wu, Edunov, Chen, and Yih}]{emnlp/KarpukhinOMLWEC20/DPR}
Vladimir Karpukhin, Barlas Oguz, Sewon Min, Patrick Lewis, Ledell Wu, Sergey Edunov, Danqi Chen, and Wen{-}tau Yih. 2020.
\newblock \href {https://doi.org/10.18653/v1/2020.emnlp-main.550} {Dense passage retrieval for open-domain question answering}.
\newblock In \emph{Proceedings of the 2020 Conference on Empirical Methods in Natural Language Processing, {EMNLP} 2020, Online, November 16-20, 2020}, pages 6769--6781. Association for Computational Linguistics.

\bibitem[{Liang et~al.(2023)Liang, Bommasani, Lee, Tsipras, Soylu, Yasunaga, Zhang, Narayanan, Wu, Kumar, Newman, Yuan, Yan, Zhang, Cosgrove, Manning, R{\'{e}}, Acosta{-}Navas, Hudson, Zelikman, Durmus, Ladhak, Rong, Ren, Yao, Wang, Santhanam, Orr, Zheng, Y{\"{u}}ksekg{\"{o}}n{\"{u}}l, Suzgun, Kim, Guha, Chatterji, Khattab, Henderson, Huang, Chi, Xie, Santurkar, Ganguli, Hashimoto, Icard, Zhang, Chaudhary, Wang, Li, Mai, Zhang, and Koreeda}]{tmlr/LiangBLTSYZNWKN23/UNSUP_HOLISTIC}
Percy Liang, Rishi Bommasani, Tony Lee, Dimitris Tsipras, Dilara Soylu, Michihiro Yasunaga, Yian Zhang, Deepak Narayanan, Yuhuai Wu, Ananya Kumar, Benjamin Newman, Binhang Yuan, Bobby Yan, Ce~Zhang, Christian Cosgrove, Christopher~D. Manning, Christopher R{\'{e}}, Diana Acosta{-}Navas, Drew~A. Hudson, Eric Zelikman, Esin Durmus, Faisal Ladhak, Frieda Rong, Hongyu Ren, Huaxiu Yao, Jue Wang, Keshav Santhanam, Laurel~J. Orr, Lucia Zheng, Mert Y{\"{u}}ksekg{\"{o}}n{\"{u}}l, Mirac Suzgun, Nathan Kim, Neel Guha, Niladri~S. Chatterji, Omar Khattab, Peter Henderson, Qian Huang, Ryan Chi, Sang~Michael Xie, Shibani Santurkar, Surya Ganguli, Tatsunori Hashimoto, Thomas Icard, Tianyi Zhang, Vishrav Chaudhary, William Wang, Xuechen Li, Yifan Mai, Yuhui Zhang, and Yuta Koreeda. 2023.
\newblock \href {https://openreview.net/forum?id=iO4LZibEqW} {Holistic evaluation of language models}.
\newblock \emph{Trans. Mach. Learn. Res.}, 2023.

\bibitem[{Liu et~al.(2024{\natexlab{a}})Liu, Lin, Hewitt, Paranjape, Bevilacqua, Petroni, and Liang}]{tacl/LiuLHPBPL24/Lost-in-the-middle}
Nelson~F. Liu, Kevin Lin, John Hewitt, Ashwin Paranjape, Michele Bevilacqua, Fabio Petroni, and Percy Liang. 2024{\natexlab{a}}.
\newblock \href {https://doi.org/10.1162/tacl\_a\_00638} {Lost in the middle: How language models use long contexts}.
\newblock \emph{Trans. Assoc. Comput. Linguistics}, pages 157--173.

\bibitem[{Liu et~al.(2024{\natexlab{b}})Liu, Wang, Wang, and Mao}]{corr/abs-2406-14848/PE-Rank}
Qi~Liu, Bo~Wang, Nan Wang, and Jiaxin Mao. 2024{\natexlab{b}}.
\newblock \href {https://doi.org/10.48550/arXiv.2406.14848} {Leveraging passage embeddings for efficient listwise reranking with large language models}.
\newblock \emph{CoRR}.

\bibitem[{Nguyen et~al.(2016)Nguyen, Rosenberg, Song, Gao, Tiwary, Majumder, and Deng}]{/nips/NguyenRSGTMD16/MSMARCO}
Tri Nguyen, Mir Rosenberg, Xia Song, Jianfeng Gao, Saurabh Tiwary, Rangan Majumder, and Li~Deng. 2016.
\newblock \href {https://ceur-ws.org/Vol-1773/CoCoNIPS\_2016\_paper9.pdf} {{MS} {MARCO:} {A} human generated machine reading comprehension dataset}.
\newblock In \emph{Proceedings of the Workshop on Cognitive Computation: Integrating neural and symbolic approaches 2016 co-located with the 30th Annual Conference on Neural Information Processing Systems {(NIPS} 2016), Barcelona, Spain, December 9, 2016}, volume 1773 of \emph{{CEUR} Workshop Proceedings}. CEUR-WS.org.

\bibitem[{Nogueira et~al.(2019)Nogueira, Yang, Cho, and Lin}]{corr/abs-1910-14424/monoBERT}
Rodrigo Nogueira, Wei Yang, Kyunghyun Cho, and Jimmy Lin. 2019.
\newblock \href {http://arxiv.org/abs/1910.14424} {Multi-stage document ranking with {BERT}}.
\newblock \emph{CoRR}, abs/1910.14424.

\bibitem[{Nogueira et~al.(2020)Nogueira, Jiang, Pradeep, and Lin}]{emnlp/NogueiraJPL20/MonoT5}
Rodrigo~Frassetto Nogueira, Zhiying Jiang, Ronak Pradeep, and Jimmy Lin. 2020.
\newblock \href {https://doi.org/10.18653/v1/2020.findings-emnlp.63} {Document ranking with a pretrained sequence-to-sequence model}.
\newblock In \emph{Findings of the Association for Computational Linguistics: {EMNLP} 2020, Online Event, 16-20 November 2020}, pages 708--718.

\bibitem[{OpenAI(2023)}]{corr/abs-2303-08774/GPT-4}
OpenAI. 2023.
\newblock \href {https://doi.org/10.48550/arXiv.2303.08774} {{GPT-4} technical report}.
\newblock \emph{CoRR}, abs/2303.08774.

\bibitem[{Pradeep et~al.(2023{\natexlab{a}})Pradeep, Sharifymoghaddam, and Lin}]{corr/abs-2309-15088/RankVicuna}
Ronak Pradeep, Sahel Sharifymoghaddam, and Jimmy Lin. 2023{\natexlab{a}}.
\newblock \href {https://doi.org/10.48550/arXiv.2309.15088} {Rankvicuna: Zero-shot listwise document reranking with open-source large language models}.
\newblock \emph{CoRR}.

\bibitem[{Pradeep et~al.(2023{\natexlab{b}})Pradeep, Sharifymoghaddam, and Lin}]{corr/abs-2312-02724/RankZephyr}
Ronak Pradeep, Sahel Sharifymoghaddam, and Jimmy Lin. 2023{\natexlab{b}}.
\newblock \href {https://doi.org/10.48550/arXiv.2312.02724} {Rankzephyr: Effective and robust zero-shot listwise reranking is a breeze!}
\newblock \emph{CoRR}, abs/2312.02724.

\bibitem[{Raffel et~al.(2020)Raffel, Shazeer, Roberts, Lee, Narang, Matena, Zhou, Li, and Liu}]{jmlr/RaffelSRLNMZLL20/T5}
Colin Raffel, Noam Shazeer, Adam Roberts, Katherine Lee, Sharan Narang, Michael Matena, Yanqi Zhou, Wei Li, and Peter~J. Liu. 2020.
\newblock \href {https://jmlr.org/papers/v21/20-074.html} {Exploring the limits of transfer learning with a unified text-to-text transformer}.
\newblock \emph{J. Mach. Learn. Res.}, 21:140:1--140:67.

\bibitem[{Ranasinghe et~al.(2021)Ranasinghe, Naseer, Hayat, Khan, and Khan}]{iccv/RanasingheNH0K21/orthogonal}
Kanchana Ranasinghe, Muzammal Naseer, Munawar Hayat, Salman~H. Khan, and Fahad~Shahbaz Khan. 2021.
\newblock \href {https://doi.org/10.1109/ICCV48922.2021.01211} {Orthogonal projection loss}.
\newblock In \emph{2021 {IEEE/CVF} International Conference on Computer Vision, {ICCV} 2021, Montreal, QC, Canada, October 10-17, 2021}, pages 12313--12323. {IEEE}.

\bibitem[{Reddy et~al.(2024)Reddy, Doo, Xu, Sultan, Swain, Sil, and Ji}]{emnlp/ReddyDXSSSJ24/FIRST}
Revanth~Gangi Reddy, JaeHyeok Doo, Yifei Xu, Md.~Arafat Sultan, Deevya Swain, Avirup Sil, and Heng Ji. 2024.
\newblock \href {https://aclanthology.org/2024.emnlp-main.491} {{FIRST:} faster improved listwise reranking with single token decoding}.
\newblock In \emph{Proceedings of the 2024 Conference on Empirical Methods in Natural Language Processing, {EMNLP} 2024, Miami, FL, USA, November 12-16, 2024}, pages 8642--8652. Association for Computational Linguistics.

\bibitem[{Robertson et~al.(1994)Robertson, Walker, Jones, Hancock{-}Beaulieu, and Gatford}]{trec/RobertsonWJHG94/BM25}
Stephen~E. Robertson, Steve Walker, Susan Jones, Micheline Hancock{-}Beaulieu, and Mike Gatford. 1994.
\newblock \href {http://trec.nist.gov/pubs/trec3/papers/city.ps.gz} {Okapi at {TREC-3}}.
\newblock In \emph{Proceedings of The Third Text REtrieval Conference, {TREC} 1994, Gaithersburg, Maryland, USA, November 2-4, 1994}, volume 500-225 of \emph{{NIST} Special Publication}, pages 109--126. National Institute of Standards and Technology {(NIST)}.

\bibitem[{Sachan et~al.(2022)Sachan, Lewis, Joshi, Aghajanyan, Yih, Pineau, and Zettlemoyer}]{emnlp/SachanLJAYPZ22/UNSUP_UPR}
Devendra~Singh Sachan, Mike Lewis, Mandar Joshi, Armen Aghajanyan, Wen{-}tau Yih, Joelle Pineau, and Luke Zettlemoyer. 2022.
\newblock \href {https://doi.org/10.18653/v1/2022.emnlp-main.249} {Improving passage retrieval with zero-shot question generation}.
\newblock In \emph{Proceedings of the 2022 Conference on Empirical Methods in Natural Language Processing, {EMNLP} 2022, Abu Dhabi, United Arab Emirates, December 7-11, 2022}, pages 3781--3797. Association for Computational Linguistics.

\bibitem[{Santhanam et~al.(2022)Santhanam, Khattab, Saad{-}Falcon, Potts, and Zaharia}]{naacl/SanthanamKSPZ22/colbertv2}
Keshav Santhanam, Omar Khattab, Jon Saad{-}Falcon, Christopher Potts, and Matei Zaharia. 2022.
\newblock \href {https://doi.org/10.18653/v1/2022.naacl-main.272} {Colbertv2: Effective and efficient retrieval via lightweight late interaction}.
\newblock In \emph{Proceedings of the 2022 Conference of the North American Chapter of the Association for Computational Linguistics: Human Language Technologies, {NAACL} 2022, Seattle, WA, United States, July 10-15, 2022}, pages 3715--3734. Association for Computational Linguistics.

\bibitem[{Schlatt et~al.(2025)Schlatt, Fr{\"{o}}be, Scells, Zhuang, Koopman, Zuccon, Stein, Potthast, and Hagen}]{ecir/SchlattFSZKZSPH25a/RDLLM}
Ferdinand Schlatt, Maik Fr{\"{o}}be, Harrisen Scells, Shengyao Zhuang, Bevan Koopman, Guido Zuccon, Benno Stein, Martin Potthast, and Matthias Hagen. 2025.
\newblock \href {https://doi.org/10.1007/978-3-031-88714-7\_31} {Rank-distillm: Closing the effectiveness gap between cross-encoders and llms for passage re-ranking}.
\newblock In \emph{Advances in Information Retrieval - 47th European Conference on Information Retrieval, {ECIR} 2025, Lucca, Italy, April 6-10, 2025, Proceedings, Part {III}}, volume 15574 of \emph{Lecture Notes in Computer Science}, pages 323--334. Springer.

\bibitem[{Sun et~al.(2023)Sun, Yan, Ma, Wang, Ren, Chen, Yin, and Ren}]{emnlp/0001YMWRCYR23/RankGPT}
Weiwei Sun, Lingyong Yan, Xinyu Ma, Shuaiqiang Wang, Pengjie Ren, Zhumin Chen, Dawei Yin, and Zhaochun Ren. 2023.
\newblock \href {https://doi.org/10.18653/v1/2023.emnlp-main.923} {Is chatgpt good at search? investigating large language models as re-ranking agents}.
\newblock In \emph{Proceedings of the 2023 Conference on Empirical Methods in Natural Language Processing, {EMNLP} 2023, Singapore, December 6-10, 2023}, pages 14918--14937.

\bibitem[{Thakur et~al.(2021)Thakur, Reimers, R{\"{u}}ckl{\'{e}}, Srivastava, and Gurevych}]{corr/abs-2104-08663/BEIR}
Nandan Thakur, Nils Reimers, Andreas R{\"{u}}ckl{\'{e}}, Abhishek Srivastava, and Iryna Gurevych. 2021.
\newblock \href {https://arxiv.org/abs/2104.08663} {{BEIR:} {A} heterogenous benchmark for zero-shot evaluation of information retrieval models}.
\newblock \emph{CoRR}.

\bibitem[{Xian et~al.(2023)Xian, Zhuang, Qin, Zamani, Lu, Ma, Hui, Zhao, Wang, and Bendersky}]{nips/listbetterpoint}
Ruicheng Xian, Honglei Zhuang, Zhen Qin, Hamed Zamani, Jing Lu, Ji~Ma, Kai Hui, Han Zhao, Xuanhui Wang, and Michael Bendersky. 2023.
\newblock \href {http://papers.nips.cc/paper\_files/paper/2023/hash/cc473bb3ec4176a5e640c3a6b5fb5239-Abstract-Conference.html} {Learning list-level domain-invariant representations for ranking}.
\newblock In \emph{Advances in Neural Information Processing Systems 36: Annual Conference on Neural Information Processing Systems 2023, NeurIPS 2023, New Orleans, LA, USA, December 10 - 16, 2023}.

\bibitem[{Yoon et~al.(2024)Yoon, Choi, Kim, Yun, Kim, and Hwang}]{acl/YoonCKYKH24/ListT5}
Soyoung Yoon, Eunbi Choi, Jiyeon Kim, Hyeongu Yun, Yireun Kim, and Seung{-}won Hwang. 2024.
\newblock \href {https://doi.org/10.18653/v1/2024.acl-long.125} {Listt5: Listwise reranking with fusion-in-decoder improves zero-shot retrieval}.
\newblock In \emph{Proceedings of the 62nd Annual Meeting of the Association for Computational Linguistics (Volume 1: Long Papers), {ACL} 2024, Bangkok, Thailand, August 11-16, 2024}, pages 2287--2308.

\bibitem[{Zheng et~al.(2024)Zheng, Zhou, Meng, Zhou, and Huang}]{iclr/Zheng0M0H24/NotRobustMC}
Chujie Zheng, Hao Zhou, Fandong Meng, Jie Zhou, and Minlie Huang. 2024.
\newblock \href {https://openreview.net/forum?id=shr9PXz7T0} {Large language models are not robust multiple choice selectors}.
\newblock In \emph{The Twelfth International Conference on Learning Representations, {ICLR} 2024, Vienna, Austria, May 7-11, 2024}. OpenReview.net.

\bibitem[{Zhu et~al.(2023)Zhu, Yuan, Wang, Liu, Liu, Deng, Dou, and Wen}]{corr/abs-2308-07107/SURVEY_LLM4IR}
Yutao Zhu, Huaying Yuan, Shuting Wang, Jiongnan Liu, Wenhan Liu, Chenlong Deng, Zhicheng Dou, and Ji{-}Rong Wen. 2023.
\newblock \href {https://doi.org/10.48550/arXiv.2308.07107} {Large language models for information retrieval: {A} survey}.
\newblock \emph{CoRR}, abs/2308.07107.

\bibitem[{Zhuang et~al.(2024)Zhuang, Qin, Hui, Wu, Yan, Wang, and Bendersky}]{naacl/ZhuangQHWYWB24/UNSUP_Beyond_YN}
Honglei Zhuang, Zhen Qin, Kai Hui, Junru Wu, Le~Yan, Xuanhui Wang, and Michael Bendersky. 2024.
\newblock \href {https://doi.org/10.18653/v1/2024.naacl-short.31} {Beyond yes and no: Improving zero-shot {LLM} rankers via scoring fine-grained relevance labels}.
\newblock In \emph{Proceedings of the 2024 Conference of the North American Chapter of the Association for Computational Linguistics: Human Language Technologies: Short Papers, {NAACL} 2024, Mexico City, Mexico, June 16-21, 2024}, pages 358--370. Association for Computational Linguistics.

\bibitem[{Zhuang et~al.(2023{\natexlab{a}})Zhuang, Qin, Jagerman, Hui, Ma, Lu, Ni, Wang, and Bendersky}]{sigir/Zhuang0J0MLNWB23/RankT5}
Honglei Zhuang, Zhen Qin, Rolf Jagerman, Kai Hui, Ji~Ma, Jing Lu, Jianmo Ni, Xuanhui Wang, and Michael Bendersky. 2023{\natexlab{a}}.
\newblock \href {https://doi.org/10.1145/3539618.3592047} {Rankt5: Fine-tuning {T5} for text ranking with ranking losses}.
\newblock In \emph{Proceedings of the 46th International {ACM} {SIGIR} Conference on Research and Development in Information Retrieval, {SIGIR} 2023, Taipei, Taiwan, July 23-27, 2023}, pages 2308--2313.

\bibitem[{Zhuang et~al.(2023{\natexlab{b}})Zhuang, Liu, Koopman, and Zuccon}]{emnlp/Zhuang0KZ23/UNSUP_QLM}
Shengyao Zhuang, Bing Liu, Bevan Koopman, and Guido Zuccon. 2023{\natexlab{b}}.
\newblock \href {https://doi.org/10.18653/v1/2023.findings-emnlp.590} {Open-source large language models are strong zero-shot query likelihood models for document ranking}.
\newblock In \emph{Findings of the Association for Computational Linguistics: {EMNLP} 2023, Singapore, December 6-10, 2023}, pages 8807--8817. Association for Computational Linguistics.

\end{thebibliography}
\newpage
\appendix
% \section{Appendix}
% \section{Additional Related Work}
% ICAE~\cite{iclr/00010WWCW24/ICAE} compresses long contexts into a compact set of learnable memory slots, enabling efficient attention over large inputs. FiD-Light~\cite{sigir/HofstatterC0Z23/fid-light} builds on the FiD architecture which improves efficiency by using only the first k token embeddings of each passage as decoder input, rather than the full context. Both approaches share the goal of reducing computational cost while retaining essential information, aligning with our use of compressed token representations for efficient reranking.

\section{Implementation Details}
% \subsection{Training Hyperparamters} \label{sec:Hyperparameter}
% Our training code is implemented based on the PyTorch framework, and we utilize DeepSpeed Stage 2 for optimization. The detailed hyperparameter configurations are provided in Table~\ref{tab:training_config}. Training took approximately 5 hours on 2 × NVIDIA RTX 3090 GPUs for T5-base, and around 40 hours on 2 × NVIDIA A6000 GPUs for T5-3B.
% \input{Tables/tab_MVP_hyperparameters}

\subsection{Passage Length Configuration} \label{sec:max_input_length}
During inference, we follow the passage length configuration from ListT5~\cite{acl/YoonCKYKH24/ListT5}, where the maximum passage length for each dataset is selected from [256, 512, and 1024] based on the average number of tokens in the query-passage pair. For the \textit{signal} dataset, however, we use a smaller maximum length of 128, considering its short input length. We found that this reduced setting did not negatively impact performance.
The final maximum input lengths used for each dataset are summarized as follows:

`dl19': 256,
`dl20': 256,
`trec-covid': 512,
`nfcorpus': 512,
`signal': 128,
`news': 1024,
`robust04': 1024,
`scifact': 512,
`touche' : 1024,
`dbpedia-entity' : 256

\section{Additional Experiments} \label{sec:additional_experiment}

\subsection{Comparison with Generation-Based Reranking}
To further validate our approach, we trained the ListT5 framework on our dataset. Following prior work~\cite{acl/YoonCKYKH24/ListT5}, the model was configured to take 5 passages as input and generate the top 2 passages. Results are shown in Table~\ref{tab:tab_MVP_generation}.

\begin{table}[]
\centering
\small  
\begin{tabular}{lc|cc}
\toprule
\textbf{Model} & \textbf{Training Data} & \textbf{DL19} & \textbf{DL20} \\
\midrule
MVP & RankDistiLLM &\textbf{74.3} & \textbf{69.2} \\
ListT5 & RankDistiLLM &72.5 & 68.5 \\
\bottomrule
\end{tabular}

\caption{nDCG@10 results comparing MVP and a ListT5 variant trained on RankDistiLLM data, using tournament sort}
\label{tab:tab_MVP_generation}
\end{table}

Despite being trained on the same dataset, our anchor-based relevance estimation with multi-view representation and reranking approach consistently outperformed the generation-based model. We attribute this performance gap to two main factors: (1) generation-based models are trained with language modeling objectives, which are not inherently aligned with ranking tasks, and (2) our method evaluates relevance from multiple perspectives and aggregates the results, enabling more accurate and robust ranking estimation.

\begin{table}[t]
\small
\centering
\renewcommand{\arraystretch}{1.2} % 행 간격 조절
\setlength{\tabcolsep}{12pt}  % 기본은 6pt 정도, 더 넓게
\begin{tabular}{l|cccc}
\toprule
\textbf{} & \textbf{5} & \textbf{10} & \textbf{20} & \textbf{100} \\
\midrule
DL19            & \textbf{74.3} & 73.7 & 74.0 & 68.1 \\
DL20            & \textbf{69.2} & 68.0 & 67.0 & 62.4 \\
\midrule
Covid           & \textbf{80.2} & 80.1 & 80.1 & 76.3 \\
NFCorpus        & \textbf{36.0} & 35.9 & 35.3 & 34.3 \\
Signal          & \textbf{32.7} & 32.1 & 31.3 & 32.0 \\
News            & \textbf{49.1} & 48.6 & 47.2 & 46.0 \\
Robust04        & 55.1 & \textbf{55.4} & 53.8 & 52.5 \\
SciFact         & \textbf{75.0} & 74.4 & 74.1 & 69.3 \\
Touche          & \textbf{39.1} & 39.0 & 36.9 & 35.2 \\
DBPedia         & 43.8 & \textbf{44.0} & 43.4 & 40.7 \\
\midrule
{BEIR Avg.} & \textbf{51.4} & 51.2 & 50.3 & 48.3 \\
\bottomrule
\end{tabular}%
\caption{nDCG@10 performance with varying candidate sampling sizes during training.}
\label{tab:sampling_comparison}
\end{table}

\subsection{Effect of Sampling Size}
% 우리는 추가적으로, 학습에 사용되는 후보 문서 수가 성능에 미치는 영향을 분석하였다. 100은 Rank-DistiLLM 원본 구성 그대로의 설정이며, 10 및 20은 각각 하나의 학습 인스턴스에 10개, 20개의 후보 문서를 샘플링하여 사용하는 구성이다. 실험 결과는 표~\ref{}에 제시되어 있다.

% 결과를 보면, 후보 문서 수가 증가할수록 모델의 성능이 오히려 감소하는 경향이 나타난다. 이는 MVP가 사용하는 ListNet 손실 함수의 특성과 관련이 있다. ListNet에서는 타깃 확률 분포 생성을 위해 softmax를 사용하는데, 후보 문서가 100개로 많아질 경우 확률 분포가 과도하게 평탄화되어 상위권 문서와 하위권 문서 간의 차별성이 약해지는 문제가 발생한다.
We further analyze the impact of the number of candidate passages used during training on model performance. The setting with 100 candidates follows the original configuration of Rank-DistiLLM, while the settings with 10 and 20 candidates involve randomly sampling 10 or 20 passages per training instance, respectively. 

% Interestingly, we observe a performance degradation as the number of candidate passages increases. We attribute this to the fact that, when the number of candidates is large, the target distribution becomes overly flattened, thereby diminishing the distinction between highly relevant and irrelevant passages.
As shown in Table~\ref{tab:sampling_comparison}, we observe a performance degradation as the number of candidate passages increases. We attribute this to two main factors. First, as described in Section~\ref{sec:ranking objective}, we adopt the ListNet loss, where the target distribution is constructed by applying a softmax over the inverse rank. Increasing the number of candidates makes this distribution overly uniform, making it harder for the model to distinguish between relevant and non-relevant passages and thereby weakening the ranking signal. Second, using fewer candidates allows us to generate more diverse combinations of passages through random sampling which exposes the model to a wider range of ranking scenarios.
% \subsection{Inference Straegy Analysis}
% \subsubsection{Impact of Sorting Strategies at Inference}
% 제안하는 모델은 Soft Prompting 방식을 사용함으로써 디코더에 입력되는 Key와 Value의 차원을 크게 줄였으며, 이를 통해 한 번에 100개의 문서를 효율적으로 평가할 수 있다. 우리는 기존 Listwise 기반 재순위화 모델에서 사용된 Sorting Algorithm들을 동일하게 적용하여 비교 실험을 수행하였다. Tournament Sort의 경우 window size는 5, output passage 수는 2로 설정하였으며, Sliding Window 방식에서는 window size 20, step size 10을 사용하였다. 관련 실험 결과는 표~\ref{}에 제시되어 있다.

% 실험 결과, 전체 후보 문서를 한번에 입력하여 평가하는 경우와, Sorting Algorithm을 사용하는 경우 성능 차이가 거의 발생하지 않았으며, 
% 실험 결과, 모든 설정에서 전체 candidate passage를 한 번에 입력하여 Anchor Vector를 생성하는 방식이 가장 높은 성능을 나타냈다. 이는 Anchor Vector 생성 시, 가능한 한 다양한 문서를 고려하는 것이 보다 정밀한 관련도 추정에 도움이 된다는 점을 시사한다.

\section{Analysis of External Biases} \label{sec:bias analysis}
In this section, we analyze the factors underlying MVP’s robustness to external biases, focusing on position and selection biases.  

\setlength{\tabcolsep}{6pt}  % 열 간격 줄이기
\renewcommand{\arraystretch}{0.95}
\begin{table}[t]\small
\centering
\footnotesize
\begin{tabular}{l|cccc}
\toprule
\textbf{Candidate Order} & \textbf{DL19} & \textbf{DL20} & \textbf{News} & \textbf{Average} \\
\midrule
\multicolumn{5}{l}{{MVP}} \\
\midrule
\quad BM25        & 74.3 & 69.2 & 49.1 & 64.2 \\
\quad Shuf.\ BM25 & 74.3 & 69.2 & 49.1 & 64.2 {\scriptsize (±0.0)} \\
\quad Rev.\ BM25  & 74.3 & 69.2 & 49.1 & 64.2 {\scriptsize (±0.0)} \\
\midrule
\multicolumn{5}{l}{ListT5 (\textit{ts}: m=5, r=2)} \\
\midrule
\quad BM25        & 71.8 & 68.1 & 48.5 & 62.8 \\
\quad Shuf.\ BM25 & 71.2 & 68.2 & 48.6 & 62.7 {\scriptsize (–0.1)} \\
\quad Rev.\ BM25  & 71.2 & 67.8 & 48.5 & 62.5 {\scriptsize (–0.3)} \\
\midrule
\multicolumn{5}{l}{RankZephyr (\textit{sw}: w=20, s=10)} \\
\midrule
\quad BM25        & 73.1 & 70.8 & 52.5 & 65.5 \\
\quad Shuf.\ BM25 & 73.1 & 70.7 & 51.3 & 65.0 {\scriptsize (–0.4)} \\
\quad Rev.\ BM25  & 72.1 & 71.5 & 51.8 & 65.1 {\scriptsize (–0.3)} \\
\midrule
\multicolumn{5}{l}{{FIRST (\textit{sw}: w=20, s=10)}} \\
\midrule
\quad BM25        & 72.4 & 71.1 & 52.4 & 65.3 \\
\quad Shuf.\ BM25 & 70.0 & 69.4 & 47.3 & 62.2 {\scriptsize (–3.1)} \\
\quad Rev.\ BM25  & 67.5 & 68.3 & 42.4 & 59.4 {\scriptsize (–5.9)} \\
\midrule
\multicolumn{5}{l}{{PE-Rank (\textit{sw}: w=20, s=10)}} \\
\midrule
\quad BM25        & 70.8 & 65.4 & 52.3 & 62.8 \\
\quad Shuf.\ BM25 & 66.0 & 58.5 & 46.8 & 57.1 {\scriptsize (–5.7)} \\
\quad Rev.\ BM25  & 67.5 & 59.1 & 46.5 & 57.7 {\scriptsize (–5.1)} \\
\bottomrule
\end{tabular}
\caption{nDCG@10 results under different candidate orders. Values in parentheses indicate change relative to the BM25 ranking order. \textit{ts} denotes tournament sort and \textit{sw} denotes sliding window. For each sorting algorithm, the basic operating unit ($m\to r$), window size ($w$), and stride ($s$) are set according to prior work.}
\label{tab:position}
%\vskip -0.1in
\end{table}

\setlength{\tabcolsep}{6pt}  % 기본값으로 복원

% \multicolumn{5}{l}{\textbf{RankVicuna-7B \textit{sw} (w=20, s=10)}} \\
% \midrule
% \quad BM25        & 66.5 & 66.4 & 45.0 & 59.3 \\
% \quad Shuf.\ BM25 & 66.9 & 63.9 & 45.3 & 58.7 {\scriptsize (–0.6)} \\
% \quad Rev.\ BM25  & 64.2 & 63.8 & 42.1 & 56.7 {\scriptsize (–2.6)} \\
% \midrule
\subsection{Robustness to Position Bias}
Position bias denotes the dependence of reranking performance on the initial candidate order. This issue typically arises when passages receive different positional embeddings within a listwise prompt. However, as shown in Table~\ref{tab:position}, which extends the candidate permutation results of Table~\ref{tab:passage_order} with additional rerankers, MVP consistently achieves the same reranking performance regardless of the initial order. This invariance arises from the encoding and decoding mechanisms of MVP.
% \subsubsection{Encoding Stage}

\noindent\textbf{Encoding Stage}
As described in Section~\ref{sec:obtaining QPE}, we prepend identical $m$ view tokens to each document $c_i$. Using the FiD architecture, each passage is then independently encoded, producing $m$ relevance vectors. Importantly, the $k$-th view token $\langle\text{v}_k\rangle$ consistently receives the same positional embedding vector $p_k$ across all query-passage pairs.
This encoding ensures that the resulting relevance vectors remain independent of the initial passage order.

\noindent\textbf{Decoding Stage}
To generate the anchor vector, the decoder receives only a single [\text{BOS}] token as input. For cross-attention, the keys and values are the relevance vectors $\{e_{1k}, e_{2k}, \dots, e_{nk}\}$ produced at the encoding stage for the $k$-th view token $\langle\text{v}_k\rangle$ (see Section~\ref{sec:anchor reranking}). Since no additional positional embedding is applied to these keys and values, the resulting anchor vector remains invariant to permutations of the relevance vectors.

Consequently, reranking is performed by measuring similarity between a relevance vector and its anchor vector. Through this mechanism, MVP achieves consistent reranking performance regardless of the initial passage order.
\subsection{Robustness to Selection Bias}

Selection bias refers to the bias inherent in the identifier tokens used to represent passages. We investigate this issue through two sets of experiments.
\subsubsection{Designs for View Tokens} \label{appendix:design_relevance_token}
% \begin{table}[t]
% \centering
% \small
% \begin{tabular}{lccc}
% \toprule
% \textbf{Rel Tokens} & \textbf{DL19} & \textbf{DL20} & \textbf{BEIR Avg.} \\
% \midrule
% SPEAR         & 74.3 & 69.2 & 51.4 \\
% \midrule
% First 4 Tokens             & 74.2 & 68.7 & 51.0 \\
% Numeric Tokens        & 73.4 & 67.7 & 50.8 \\
% Alphabetic Tokens       & 73.2 & 68.1 & 51.1 \\
% \bottomrule
% \end{tabular}
% \caption{Comparison of different relevance token designs}
% \label{tab:special_tokens}
% \vskip -0.1in
% \end{table}

% \setlength{\tabcolsep}{3pt}
\begin{table}[t]
\renewcommand{\arraystretch}{1.2} % 행 간격 조절
\centering
% \footnotesize
\resizebox{\linewidth}{!}{ % 가로 길이 자동 조절
\small
\begin{tabular}{l|cccc}
\toprule
\textbf{} & \textbf{EXTRA ID} & \textbf{FIRST 4} & \textbf{Numeric} & \textbf{Alphabetic} \\
\midrule
DL19         & \textbf{74.3} & 74.2 & 73.4 & 73.2 \\
DL20         & \textbf{69.2} & 68.7 & 67.7 & 68.1 \\
\midrule
Covid        & \textbf{80.2} & 78.4 & 78.5 & 78.8 \\
NFCorpus     & \textbf{36.0} & 35.7 & 35.5 & 35.3 \\
Signal       & \textbf{33.0} & 32.2 & 32.0 & 33.0 \\
News         & 49.1 & \textbf{49.4} & 48.7 & 49.0 \\
Robust04     & \textbf{55.1} & 54.2 & 54.1 & 54.9 \\
SciFact      & \textbf{75.0} & 74.6 & 73.5 & 73.8 \\
Touche       & 39.1 & 39.7 & \textbf{40.4} & \textbf{40.4} \\
DBPedia      & 43.8 & \textbf{44.3} & 43.9 & 43.8 \\
\midrule
{BEIR Avg.} & \textbf{51.4} & 51.0 & 50.8 & 51.1 \\
\bottomrule
\end{tabular}
}
\caption{nDCG@10 results for different view token designs.}
\label{tab:special_token_design}
\end{table}

\setlength{\tabcolsep}{6pt}  % 기본값으로 복원

To analyze the impact of view token design on re-ranking performance, we compare three alternative configurations: 
(1) First 4 Tokens: Following the FiD-Light~\cite{sigir/HofstatterC0Z23/fid-light} approach, the first four tokens in the input prompt are reused without introducing dedicated special tokens; 
(2) Numeric Tokens: View tokens are replaced with number-based tokens (1, 2, 3, 4); and
(3) Alphabetic Tokens: Character-based tokens (A, B, C, D) are used as view tokens.

Table~\ref{tab:special_token_design} shows that the \texttt{<extra\_id>} tokens from T5 tokenizer, as adopted in MVP, yields the best performance. This result suggests that: (1) Learnable token embeddings specifically trained to encode query-passage relevance are more effective than simply reusing the first prompt tokens. (2) Moreover, numeric and alphabetic identifiers may already carry semantic meaning from pretraining, leading to potential conflicts with their intended function as compression tokens, ultimately resulting in degraded performance.

\subsubsection{Identifier Reordering} 
Existing generation-based listwise rerankers rely on identifier tokens to produce outputs, which can introduce selection bias. To further examine this issue beyond the experiments in Section~\ref{sec:robustness}, we conducted additional evaluations on various listwise rerankers. The results are summarized in Table~\ref{tab:selection}.

The results confirm that models using numeric or alphabetic identifiers are sensitive to identifier reordering, with most models exhibiting performance drops. Even in ListT5, which leverages the FiD architecture, we observe minor performance variations. In contrast, MVP avoids this issue by employing randomly initialized view tokens shared across passages and computing relevance scores directly from passage-specific vectors.
\setlength{\tabcolsep}{6pt}  % 열 간격 줄이기
\renewcommand{\arraystretch}{0.95}
\begin{table}[t]
\centering
\footnotesize
\begin{tabular}{l|cccc}
\toprule
\textbf{Identifier Order} & \textbf{DL19} & \textbf{DL20}  & \textbf{News}& \textbf{Average} \\
\midrule
\multicolumn{5}{l}{MVP} \\
\midrule
\quad - & 74.3 & 69.2 & 49.1 & 64.2 \\
\midrule
\multicolumn{5}{l}{ListT5 (\textit{ts}: m=5, r=2)} \\
\midrule
\quad Original        & 71.8 & 68.1 & 48.5 & 62.8 \\
\quad Shuffle & 71.4 & 68.3 & 49.2 & 63.0 {\scriptsize (+0.2)} \\
\quad Reverse & 71.4 & 67.8 & 49.4 & 62.9 {\scriptsize (+0.1)} \\
\midrule
\multicolumn{5}{l}{RankZephyr (\textit{sw}: w=20, s=10)} \\
\midrule
\quad Original        & 73.1 & 70.8 & 52.5 & 65.5 \\
\quad Shuffle & 71.3 & 67.3 & 46.7 & 61.8 {\scriptsize (–3.7)} \\
\quad Reverse & 69.3 & 63.9 & 47.2 & 60.1 {\scriptsize (–5.3)} \\
\midrule
\multicolumn{5}{l}{{FIRST (\textit{sw}: w=20, s=10)}} \\
\midrule
\quad Original        & 72.4 & 71.1 & 52.4 & 65.3 \\
\quad Shuffle & 71.2 & 69.2 & 49.1 & 63.2 {\scriptsize (–2.1)} \\
\quad Reverse & 71.0 & 68.2 & 48.5 & 62.6 {\scriptsize (–2.7)} \\
\midrule
\multicolumn{5}{l}{{PE-Rank (\textit{sw}: w=20, s=10)}} \\
\midrule
\quad Original        & 70.8 & 65.4 & 52.3 & 62.8 \\
\quad Shuffle & 70.5 & 65.3 & 51.9 & 62.6 {\scriptsize (–0.2)} \\
\quad Reverse  & 70.3 & 65.3 & 52.2 & 62.6 {\scriptsize (–0.2)} \\
\bottomrule
\end{tabular}
\caption{nDCG@10 results under different identifier orders. Values in parentheses indicate change relative to the original identifier configuration. \textit{ts} denotes tournament sort and \textit{sw} denotes sliding window.
% \textit{ts} denotes tournament sort and \textit{sw} denotes sliding window. For each sorting algorithm, the basic operating unit ($m\to r$), window size ($w$), and stride ($s$) are set according to prior work.
}
\label{tab:selection}
%\vskip -0.1in
\end{table}

\setlength{\tabcolsep}{6pt}  % 기본값으로 복원

% \multicolumn{5}{l}{\textbf{RankVicuna-7B \textit{sw} (w=20, s=10)}} \\
% \midrule
% \quad BM25        & 66.5 & 66.4 & 45.0 & 59.3 \\
% \quad Shuf.\ BM25 & 66.9 & 63.9 & 45.3 & 58.7 {\scriptsize (–0.6)} \\
% \quad Rev.\ BM25  & 64.2 & 63.8 & 42.1 & 56.7 {\scriptsize (–2.6)} \\
% \midrule

\section{Additional Analysis of MVP}
\subsection{Impact of Orthogonal Regularization} \label{sec:view similiarity}
% \begin{table}[t]
% \centering
% \small
% \begin{tabular}{l|cc|cc}
% \toprule
% \multirow{2}{*}{} & \multicolumn{2}{c|}{\textbf{Relevance Vectors}} & \multicolumn{2}{c}{\textbf{Anchor Vectors}} \\
% \cmidrule(lr){2-3} \cmidrule(lr){4-5}
%                  & \textbf{Mean} & \textbf{Std} & \textbf{Mean} & \textbf{Std} \\
% \midrule
% {MVP}           & 0.8815 & 0.0232 & 0.9800 & 0.0062 \\
% {w/o Orthogonal}  & 0.4910 & 0.0229 & -0.0025 & 0.0010 \\
% \bottomrule
% \end{tabular}
% \caption{Mean and standard deviation of cosine similarities among relevance and anchor vectors.}
% \label{tab:vector_similarity}
% \end{table}

\begin{table}[t]
\centering
\small
\begin{tabular}{l|c|c}
\toprule
\textbf{} & \textbf{Relevance Vectors} & \textbf{Anchor Vectors}\\
\midrule
{MVP}           & 0.4910 (0.0229) & -0.0025 (0.0010) \\
{w/o Orthogonal}  & 0.8815 (0.0232) & 0.9800 (0.0062) \\
\bottomrule
\end{tabular}
\caption{Mean (standard deviation) of pairwise cosine similarities. Similarities are calculated respectively among relevance vectors and anchor vectors.}
\label{tab:vector_similarity}
\end{table}

\FloatBarrier
\begin{table}[!t]
\renewcommand{\arraystretch}{1.2}
\small
\centering
\setlength{\tabcolsep}{2pt}  % 추가된 부분

\begin{tabular}{l|cccccc}
\toprule
\textbf{} & \textbf{View 1} & \textbf{View 2} & \textbf{View 3} & \textbf{View 4} & \textbf{MAX} & \textbf{Mean} \\
\midrule
DL19       & 72.8 & 71.5 & 72.7 & 73.5 & 73.5 & \textbf{74.3} \\
DL20       & 66.3 & 65.9 & 68.0 & 68.3 & 68.4 & \textbf{69.2} \\
\midrule
Covid      & 78.9 & 78.8 & 79.9 & 80.0 & 80.1 & \textbf{80.2} \\
NFCorpus   & 35.7 & 32.4 & 31.4 & \textbf{36.2} & 33.1 & 36.0 \\
Signal     & 30.6 & \textbf{33.1} & 32.6 & 32.7 & 33.4 & 32.7 \\
News       & 47.5 & 44.7 & \textbf{49.9} & 48.0 & 46.9 & 49.1 \\
Robust04   & 52.2 & 51.4 & 54.4 & \textbf{55.3} & 53.4 & 55.1 \\
SciFact    & 74.1 & 58.5 & 57.0 & 74.7 & 73.4 & \textbf{75.0} \\
Touche     & 34.2 & 37.2 & 38.2 & 38.6 & 37.9 & \textbf{39.1} \\
DBPedia    & 43.1 & 42.0 & 41.9 & 43.6 & 43.7 & \textbf{43.8} \\
\midrule
{BEIR Avg.}  & 49.5 & 47.2 & 48.2 & 51.1 & 50.2 & \textbf{51.4} \\
\bottomrule
\end{tabular}
\caption{nDCG@10 comparison of view-wise score aggregation methods, including individual views, \textit{MAX}, and \textit{Mean}. The \textit{Mean} strategy corresponds to the default aggregation method used in our proposed framework  MVP.}
\label{tab:view_aggregation}
\end{table}

To verify whether orthogonality promotes separation across views, we analyze the pairwise cosine similarities within anchor vectors and relevance vectors on the DL20 dataset, which contains 54 queries, each associated with 100 candidate passages. We compare the results between MVP and its variant without orthogonality regularization. For anchor vectors, we compute the average pairwise similarities among the 4 anchors for each query\footnote{Four vectors generate six unique pairs.}
 and report the average across 54 queries. For relevance vectors, we also compute the average pairwise similarities among the four vectors produced for each query-passage pair, and report the average over all 5,400 pairs.
 
The results are presented in Table~\ref{tab:vector_similarity}. As shown, removing the orthogonality constraint leads to a substantial increase in similarity among both anchor and relevance vectors. This indicates that the relevance vectors capture highly similar signals, and the anchor vectors assess relevance using overlapping criteria. Consequently, this reduces view diversity and leads to performance degradation.

\subsection{Effectiveness of View Aggregation}

We conducted an additional analysis to verify whether the proposed model effectively aggregates information from each view. Table~\ref{tab:view_aggregation} presents the results of this analysis, where each column represents a different aggregation strategy. Specifically, columns labeled \textit{View 1}, ..., \textit{View 3} show performance when reranking is performed using scores from each individual view alone, while the column labeled \textit{Max} indicates performance obtained by selecting the highest relevance score among all views as the final relevance score. Lastly, the column labeled \textit{Mean} corresponds to our proposed MVP approach, where the final relevance score is calculated by averaging scores across all views.

Experimental results demonstrate that, the MVP approach of averaging scores across views consistently outperforms in most scenarios. In contrast, the \textit{MAX} strategy results in decreased performance, which can be attributed to the inconsistency introduced by selecting the final score from different views. Since each view captures distinct relevance perspectives, relying on a single highest score may lead to instability and undermine the overall ranking consistency.

\begin{table}[t]\small
\centering
\renewcommand{\arraystretch}{1.3} % 행 간격 조절
\begin{tabular}{@{}l|ccc@{}}
\toprule
\textbf{} & \textbf{MVP} & \textbf{w/o $\mathcal{L}_{\text{Orthogonal}}$} & \makecell{\textbf{w/o Multi-view} \\ \textbf{Encoding}} \\
\midrule
Covid      & \textbf{80.2} & 79.1 & 78.8 \\
NFCorpus   & \textbf{36.0} & 35.6 & 35.8 \\
Signal     & \textbf{32.7} & 31.4 & 32.6 \\
News       & \textbf{49.1} & 48.1 & 48.7 \\
Robust04   & 55.1          & 54.6 & \textbf{55.2} \\
SciFact    & \textbf{75.0} & 74.3 & 73.2 \\
Touche     & \textbf{39.1} & 38.4 & \textbf{39.1} \\
DBPedia    & 43.8          & 43.9 & \textbf{44.2} \\
\midrule
BEIR Avg.       & \textbf{51.4} & 50.7 & 50.9 \\
\bottomrule
\end{tabular}
\caption{Full BEIR results for the ablation study on training strategies.}
\label{tab:ablation_full_beir}
\end{table}

\subsection{Full Reranking Results from Ablation Studies}
Following the analysis in Section~\ref{sec:training strategies}, Table~\ref{tab:ablation_full_beir} reports the full reranking results from the ablation experiments.

\end{document}